\documentclass[grl]{agutex}

\usepackage{graphicx}
\usepackage{amssymb}
\authorrunninghead{HELMBOLDT \& INTEMA}

\titlerunninghead{SPECTRAL MAPPING OF IONOSPHERIC WAVES}

\authoraddr{J.\ F.\ Helmboldt, US Naval Research Laboratory, 
Code 7213, 4555 Overlook Ave. SW, Washington, DC 20375 
(joe.helmboldt@nrl.navy.mil)}

\authoraddr{H.\ T.\ Intema, National Radio Astronomy Observatory, 
1003 Lopezville Rd, Socorro, NM 87801 
(hintema@nrao.edu)}

\begin{document}

\title{Advanced spectral analysis of ionospheric waves observed with sparse arrays}

\authors{J.\ F.\ Helmboldt \altaffilmark{1} and H.\ T.\ Intema \altaffilmark{2}}

\altaffiltext{1}{US Naval Research Laboratory, Washington, DC, USA.}
\altaffiltext{2}{National Radio Astronomy Observatory, Socorro, NM, USA.}

\begin{abstract}
This paper presents a case study from a single, six-hour observing period to illustrate the application of techniques developed for interferometric radio telescopes to the spectral analysis of observations of ionospheric fluctuations with sparse arrays.  We have adapted the deconvolution methods used for making high dynamic range images of cosmic sources with radio arrays to making comparably high dynamic range maps of spectral power of wavelike ionospheric phenomena.  In the example presented here, we have used observations of the total electron content (TEC) gradient derived from Very Large Array (VLA) observations of synchrotron emission from two galaxy clusters at 330 MHz as well as GPS-based TEC measurements from a sparse array of 33 receivers located within New Mexico near the VLA.  We show that these techniques provide a significant improvement in signal to noise (S/N) of detected wavelike structures by correcting for both measurement inaccuracies and wavefront distortions.  This is especially true for the GPS data when combining all available satellite/receiver pairs, which probe a larger physical area and likely have a wider variety of measurement errors than in the single-satellite case.  In this instance, we found the peak S/N of the detected waves was improved by more than an order of magnitude.  The data products generated by the deconvolution procedure also allow for a reconstruction of the fluctuations as a two-dimensional waveform/phase screen that can be used to correct for their effects.
\end{abstract}

\begin{article}

\section{Introduction}
Earth's ionosphere is host to several wave phenomena.  These are typically referred to with the umbrella term ``traveling ionospheric disturbance'' (TID), but span a wide range of sizes, speeds, and amplitudes.  There are also several mechanisms which contribute to the generation and maintenance of these waves.  Included among these is forcing from the lower atmosphere, usually via gravity waves.  These can be orographic \citep[e.g.,][]{vadas10} or acoustic \citep[e.g.,][]{hickey09} gravity waves that, in either case, perturb both the temperature and wind speed within the ionosphere, which leads to perturbations in electron density and total electron content (TEC).  There are also inherent instabilities within the ionosphere such as the Perkins instability \citep{perkins73} and a coupled instability between sporadic-E layers and the Perkins instability \citep{cosgrove04a} that favor density perturbations oriented in a specific direction.  Relatively large TIDs are typically generated by geomagnetic storms \citep{borries09} via auroral heating.  TIDs are also commonly found at dusk and dawn \citep[e.g.,][]{hernandez06}, the result of disturbances associated with recombination/ionization occurring at sunset/sunrise.\par
It is common practice to designate TIDs according to size with medium scale TIDs (MSTIDs) having wavelengths $\lesssim\!500$ km and speeds $\sim\!\!100$ m s$^{-1}$ \citep[e.g.,][]{hernandez06,shiokawa08}, and large scale TIDs (LSTIDs) having wavelengths $\sim\!\!2000$ km and speeds of $\sim\!\!700$ m s$^{-1}$ or more \citep[e.g.,][]{borries09}.  However, these classifications may be more of a reflection of instrument capabilities rather than a real, bimodal population with significant physical differences.  There may in fact be something of a continuum of ionospheric waves with various generation mechanisms.  Understanding this rich population of waves is key to understanding the dynamics of Earth's ionosphere over a wide variety of scales.\par
It is with this motivation that we have developed advanced spectral analysis techniques aimed at improving the capability of detecting wavelike ionospheric disturbances over a large span of sizes and amplitudes, even when these waves are relatively sparsely (but not under) sampled.  Part of the difficulty in detecting such waves lies in the assimilation of data from several detectors/sensors that have different biasses and that probe different parts of the wavefront.  In the presence of significant wavefront distortions, both physical and non-physical effects can cause the time series measured by each sensor to be further out of phase than they should be, effectively blurring and distorting the signature of such a wave in the spectral domain.  This is essentially identical to the problem experienced by radio astronomers when constructing images of cosmic sources from radio-frequency interferometers.  Both instrumental and atmospheric/ionospheric amplitude and phase fluctuations that vary on relatively short time scales significantly degrade the quality of these synthesis images.  Consequently, sophisticated deconvolution techniques have been developed to mitigate these effects.\par
Here, we describe how we have adapted these techniques to the spectral analysis of ionospheric fluctuations.  We present this using a case study of a single six-hour observing run on 14 September 2012.  We separately analyze contemporaneous observations of the TEC gradient made with the Very Large Array (VLA) in central New Mexico and TEC fluctuations measure with a relatively sparse array of 33 GPS receivers distributed throughout New Mexico.  Sec.\ 2 details the techniques involved, Sec.\ 3 and 4 describe their application to the VLA and GPS data, and Sec.\ 5 discusses the results of this analysis.
\section{The Method}
\subsection{Reformulating the Problem}
The spectral analysis presented in this section is designed to work with relatively sparse arrays of sensors such as radio-frequency antennas, GPS receivers, and ionosondes/sounders.  At the heart of the method is the well known construct, the discrete Fourier transform (DFT).  In order to fully deconstruct observations from a two-dimensional array into wave components, one must perform a three-dimensional DFT, one temporal dimension and two spatial, according to
%\begin{linenomath}
\begin{equation}
D(\nu,\xi_x,\xi_y) = \frac{1}{N_e N_t} \sum_{j=1}^{N_e} \sum_{k=1}^{N_t} S_{j,k} e^{-2\pi i(\nu t_k - \xi_x x_j - \xi_y y_j)}
\label{dft}
\end{equation}
%\end{linenomath}
where $N_e$ is the number of array elements, $N_t$ is the number of temporal samples used in the DFT, $S_{j,k}$ is the signal at element $j$ and time sample $k$, $x$ and $y$ are the east-west and north-south positions of the elements, respectively, $\xi_x$ and $\xi_y$ are the corresponding spatial frequencies, and $\nu$ is the temporal frequency.  Formally speaking, this is really a DFT in the time domain and a two-dimensional inverse DFT in the spatial domain.\par
Note that equation (1) assumes that the observed TEC fluctuations can be characterized as moving within a thin layer at a fixed height.  Here, we have assumed an altitude of 300 km.  For a case when the field of regard is relatively large (e.g., combining all satellites observed with one GPS array; see Sec.\ 3.2 and Sec.\ 4), the quantities $x$ and $y$ are computed as east-west and north-south arc-lengths within this thin spherical shell.  Contrast this with the approximation typically used within radio astronomy where a portion of the sky or ``celestial sphere'' is approximated with a plane to reduce the imaging problem to a two-dimensional fast Fourier Transform (FFT).  This approximation only holds when the field of view (in radians) is $\lesssim\!\sqrt{\lambda / B}$, where $\lambda$ is the observing wavelength and $B$ is the length of the longest baseline \citep{perley99}.\par
Because it is the spatial rather than the temporal sampling that is, in this case, relatively sparse, it is convenient to define a quantity $T_j(\nu)$, which is the temporal DFT for element $j$ evaluated at a temporal frequency $\nu$.  It is then straightforward to compute a two-dimensional power spectrum at each value of $\nu$.  However, one may wish to increase the signal to noise ratio (S/N) of this power spectrum by using a sliding temporal window of width $N_t$, stepped by $N_I$ time steps.  In other words, for time interval $k$, $T_{j,k}(\nu)$ can be computed as
%\begin{linenomath}
\begin{equation}
T_{j,k} (\nu) = \frac{1}{N_t} \sum_{l =k}^{N_t+k-1} S_{j,l} e^{-2 \pi i \nu t_l}
\label{cvolt}
\end{equation}
%\end{linenomath}
where the summation can be performed simultaneously for all values of $\nu$ within a single time interval using an FFT if one so chooses.  One may then compute an average power spectrum according to
%\begin{linenomath}
\begin{equation}
P(\nu,\xi_x,\xi_y) = \frac{1}{N_I} \sum_{k=1}^{N_I} \left | \frac{1}{N_e} \sum_{j=1}^{N_e} T_{j,k}(\nu) e^{2\pi i(\xi_x x_j + \xi_y y_j)} \right |^2
\label{pspec}
\end{equation}
%\end{linenomath}
\par
While it is straightforward to compute this sum over a chosen three-dimensional ($\nu$, $\xi_x$, and $\xi_y$) grid, such a sum has a significant drawback, especially for sparse arrays.  The complex sum within the vertical bars in the right hand side of equation (\ref{pspec}) yielded at each interval is convolved with a combination of the sampling function of the array and any element-based amplitude and phase errors, which may be instrumental and/or physical (i.e., localized wavefront distortions) in nature.  Because of this, the power spectrum $P$ computed according to equation (\ref{pspec}) will have significant artifacts, including sidelobes from the Fourier transform of the array sampling function, or impulse response function (IRF), and any instrumental/physical distortions.\par
The difficulty in mitigating these effects is that they are convolved with each individual complex sum, rather than the mean power spectrum itself, making the application of standard deconvolution techniques challenging.  \citet{helmboldt12c} attempted an ad hoc deconvolution of only the complex IRF (i.e., not the additional distortions) from one complex sum at a time, with some success \citep[see][for details of the deconvolution method applied]{helmboldt12c}.  However, a more thorough and robust application of advanced deconvolution techniques developed by the radio astronomy community can be developed by recasting equation (\ref{pspec}) in a form that more closely resembles what is measured by an interferometric telescope.\par
Specifically, such an interferometer computes correlations of signals between pairs of antennas, or ``baselines.''  For a small field of view, these correlations, or ``visibilities,'' are related to the distribution of intensity on the sky through an inverse Fourier transform.  By expanding and re-ordering the terms of the sums in the righthand side of equation (\ref{pspec}), we can relate the fluctuation power spectrum to the inverse DFT of a set of similar correlations.\par
First, let us express the complex value $T$ as $T=T_R+iT_I$, and let us define the complex product $T\mbox{exp}[2\pi i(\xi_x x + \xi_y y)] = T\mbox{exp}(i\phi) = R + iI$.  Using these definitions, equation (\ref{pspec}) becomes
%\begin{linenomath}
\begin{equation}
P=\frac{1}{N_e^2} \left [ \sum_{j=1}^{N_e} \left < R_j^2+I_j^2 \right > + 2 \sum_{j=1}^{N_e-1} \sum_{k=j+1}^{N_e} \left < R_j R_k + I_j I_k \right > \right ]
\label{pspec2}
\end{equation}
%\end{linenomath}
where the brackets, $\left < \right >$, denote an average over $N_I$ time steps.  Equation (\ref{pspec2}) can be further altered to the following form
%\begin{linenomath}
\begin{eqnarray}
P = \frac{1}{N_e^2} \sum_{j=1}^{N_e} \left <|T_j|^2 \right > & \nonumber \\
 + \frac{2}{N_e^2} \sum_{j=1}^{N_e-1} \sum_{k=j+1}^{N_e} & \left < T_{R,j}T_{R,k} + T_{I,j}T_{I,k} \right > \cos(\phi_i-\phi_k) \nonumber \\
 + \frac{2}{N_e^2} \sum_{j=1}^{N_e-1} \sum_{k=j+1}^{N_e} & \left < T_{R,j}T_{I,k} - T_{I,j}T_{R,k} \right > \sin(\phi_i-\phi_k)
 \label{pspec3}
 \end{eqnarray}
%\end{linenomath}
 \par
If we define a complex quantity, $C_{j,k} = \left < T_jT_k^\ast \right >$, which is analogous to a visibility measured with an interferometer, we can rewrite the power spectrum equation again as
%\begin{linenomath}
\begin{equation}
P = \frac{1}{N_e^2} \sum_{j=1}^{N_e} \left <|T_j|^2 \right > + \frac{2}{N_e^2} \sum_{j=1}^{N_e-1} \sum_{k=j+1}^{N_e} \Re \left \{ C_{j,k}e^{i(\phi_j-\phi_k)} \right \}
\label{pspec4}
\end{equation}
%\end{linenomath}
The first term in the righthand side of equation (\ref{pspec4}) basically represents the noise floor of the power spectrum.  This is of little practical use for identifying the spectral signatures of waves and can basically be ignored for this purpose.  If we define new quantities $\alpha_{j,k}=x_j-x_k$ and $\beta_{j,k}=y_j-y_k$ one can see that $\phi_j-\phi_k=2\pi(\xi_x \alpha + \xi_y \beta)$, making the power spectrum the real part of the two-dimensional inverse DFT of $C(\alpha,\beta)$.  We can also identify the parity condition used in radio interferometry, which recognizes that $C(\alpha,\beta) = C^\ast (-\alpha,-\beta)$, to obtain two measurements in the ($\alpha,\beta$) plane for every correlation, effectively eliminating the factor of two in the second term on the righthand side of equation (\ref{pspec4}).  Finally, we can incorporate any element-based amplitude and phase variations, caused by either instrumental errors or real wavefront distortions, by including a complex gain term for each element such that $\widetilde{T} = gT$ where $\widetilde{T}$ is the measured quantity and $T$ is the quantity for the unperturbed wave(s).  If this gain is stable over the integration time of $N_I$ time steps, then the observed correlations are $\widetilde{C}_{j,k}=g_jg^\ast_{k}C_{j,k}$.  The measured power spectrum is then
%\begin{linenomath}
\begin{equation}
\widetilde{P}(\nu,\xi_x,\xi_y) = \frac{1}{N_e^2} \sum_{j=1}^{N_e-1} \sum_{k=j+1}^{N_e} g_j(\nu) g_k^\ast (\nu) C_{j,k}(\nu)e^{2\pi i (\xi_x \alpha + \xi_y \beta)}
\label{final}
\end{equation}
%\end{linenomath}
where the parity condition is used to better cover the ($\alpha,\beta$) plane, eliminating the factor of two seen in previous equations, and the imaginary part of the righthand side is essentially noise and is discarded.  This is basically the same equation that relates the intensity of a cosmic source on the sky to the visibilities measured by an interferometer \citep[see, e.g.,][]{thompson91}.

\subsection{Deconvolution}
Armed with equation (\ref{final}), we can now apply standard imaging and deconvolution techniques to produce a map of the unperturbed wave spectrum as a function of $\xi_x$ and $\xi_y$ at each temporal frequency.  These techniques are described in great detail in the literature and in several standard texts such as \citet{thompson91} and \citet{taylor99}, so we will only briefly discuss them here.  The method commonly used is referred to as self-calibration.  This starts with the designation of a model for the source (here, waves) being imaged.  This model need not be of high fidelity because it will be refined within the deconvolution process.  This model can be established using any prior knowledge of the source structure.  In our case, as we do not have a practical means for estimating what ionospheric waves are expected, we use an initial power spectrum computed without any calibration applied to estimate a wave model for the first iteration.  This model is used to compute values for the visibilities/correlations, $\widehat{C}$, which are divided into the observed values, yielding the quantity
%\begin{linenomath}
\begin{equation}
X_{j,k} = \frac{\widetilde{C}_{j,k}}{\widehat{C}_{j,k}} = g_j g_k^\ast + n_{j,k}
\label{norm}
\end{equation}
%\end{linenomath}
for a baseline of elements $j$ and $k$, where $n$ is noise from (1) random noise within the data, (2) gain variations on time scales less than the integration time used to generate $\widetilde{C}$, and (3) inaccuracies within the model.  Since for $N_e$ elements, there are $N_e(N_e-1)/2$ unique baselines, one can determine the element-based complex gains with a significantly over-constrained, non-linear fitting process that minimizes the quantity
%\begin{linenomath}
\begin{equation}
\Delta = \sum_{j=1}^{N_e-1} \sum_{k=j+1}^{N_e} \left | \widehat{C}_{j,k} \right |^2 \left | X_{j,k} - g_j g_k^\ast \right |^2
\label{fit}
\end{equation}
%\end{linenomath}
Generally, for the initial iteration, the gains are set to unity, and we have adopted this practice for our application.  Thus, the procedure essentially iterates among equations (\ref{final})--(\ref{fit}) until convergence.
\par
The degree to which the self-calibration fitting procedure is over-constrained is one of its biggest strengths, but it can also be a major drawback.  If the fit is not properly constrained, it has a tendency to ``fit the noise,'' that is, to determine gain corrections that will force fluctuations within the data due to noise to conform to the chosen model.  One can avoid this problem by solving for one gain solution over a time interval that is long relative to that used for the computation of the correlations.  This forces the gain solutions to be stable over that time period, effectively mitigating the tendency to fit noisy fluctuations on time scales smaller than that interval.  However, as will become clear in Sec.\ 3 and 4, the nature of the analysis presented here precludes the use of this technique.  Instead, we have employed another method know as ``phase-only'' self-calibration.  As the name implies, this version of self-calibration only solves for the phase components of the complex gains, and not the amplitudes.  Essentially, this significantly limits the flexibility of the self-calibration fitting process, which is why phase-only self-calibration is a well-established, safe method for producing astronomical images with realistic noise characteristics \citep[see, e.g.,][]{cornwell99}.  For all of the analysis presented here, we have used phase-only self-calibration, but do not rule out the possibility of developing this technique further in the future to incorporate amplitude-and-phase self-calibration.\par
Within each iteration, before the gain solutions or phase corrections are determined and applied to the data, a new image (in our case, a two-dimensional spectrum) is made and used to establish a new model.  This is usually done using one of two methods, the CLEAN algorithm or the maximum entropy method (MEM).  CLEAN approximates the image with a series of delta functions, called ``CLEAN components," convolved with the IRF computed from the layout of the elements.  It is generally accepted to be more appropriate for images populated mostly with discrete, marginally resolved sources.  This is often the case in radio astronomy, and we will show in Sec.\ 2 that it is also the case with the power spectrum maps generated with our VLA and GPS data.\par
 In contrast, MEM utilizes a non-linear fit that maximizes the image ``entropy,'' a quantity that is maximized when the deconvolved image is positive with a compressed range in pixel values.  MEM algorithms applied to interferometer data also simultaneously minimize the $\chi^2$ difference between the measured and model visibilities inferred from the deconvovled image.  MEM generally produces smoother-looking images and performs better than CLEAN on images with a relatively large amount of extended structure.  However, for images dominated by point sources, CLEAN typically performs as well as, if not better than, MEM and is substantially faster \citep[for a more detailed discussion and comparison of both methods, see][]{cornwell99b}.  Thus, for the power spectra generated from our VLA and GPS data, which are generally lack extended structures, we have opted to employ a CLEAN-based algorithm.  However, we note that nothing about the procedure detail here explicitly precludes the use of MEM.\par
Within the CLEAN algorithm, the locations and intensities of the CLEAN components are determined iteratively from the image by identifying the pixel with the largest absolute value, subtracting a scaled version of the IRF from the image centered at that pixel, and repeating the process on the difference image until the pixel with the largest absolute value is actually negative.  Other stopping criteria may be used, but this criterion is the one most commonly employed.  The CLEAN components can then be restored to the image by convolving them with a Gaussian IRF, similar in width to the main lobe of the actual IRF, and then adding them to the final difference image.  Thus, the algorithm has effectively ``cleaned'' the effect of IRF sidelobes from the initial image.\par
\begin{figure}
\noindent\includegraphics[width=4in]{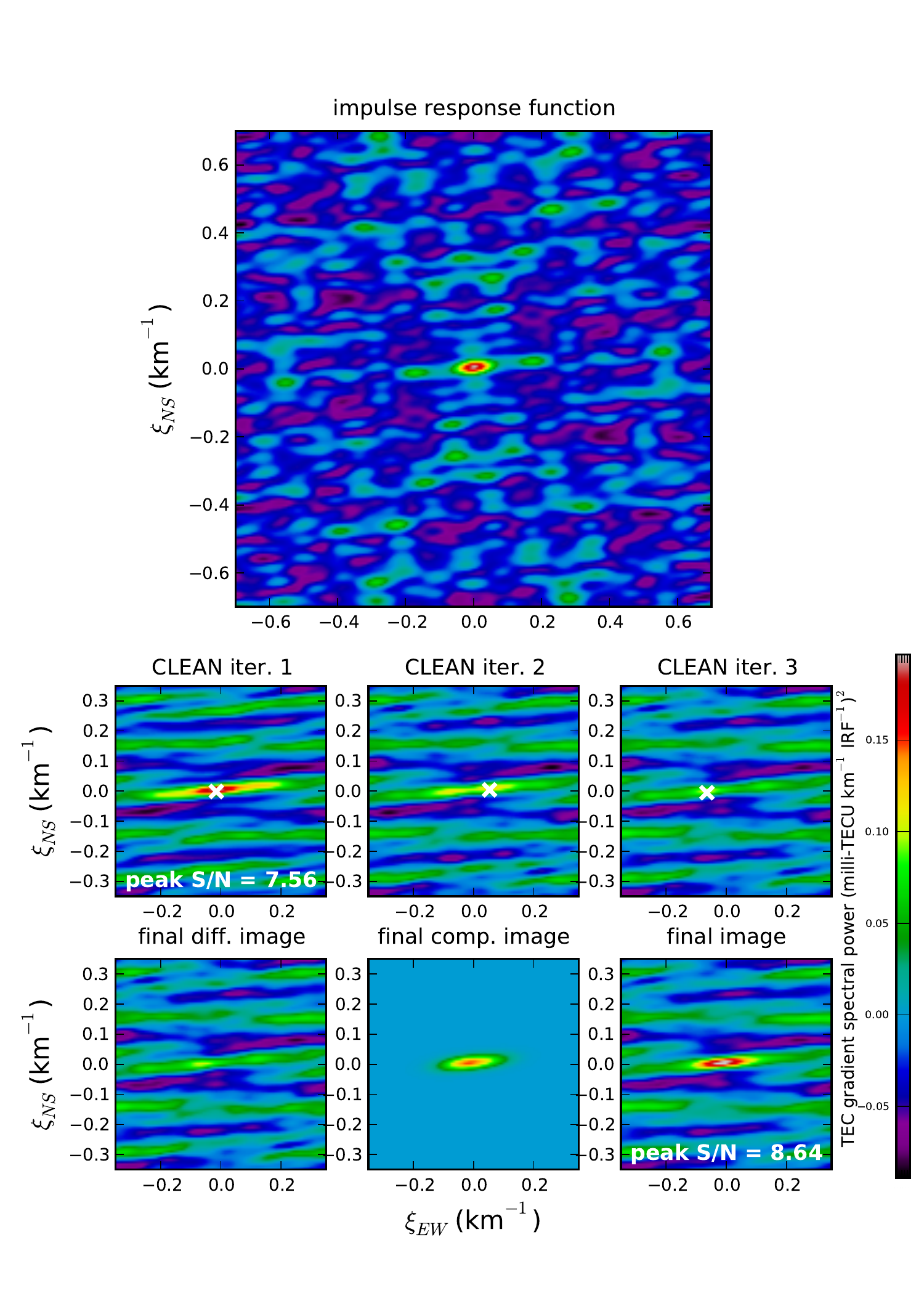}
\caption{An example of the CLEANing process as applied to a single two-dimensional fluctuation spectrum derived from VLA data (see Sec.\ 3.1 for more details regarding these data).  The largest panel shows the IRF computed using the positions of the VLA antennas.  The panels below it show three CLEAN iterations as well as the final difference image, the restored CLEAN component image, and the sum of the two as the final image.}
\label{exclean}
\end{figure}

An example of the application of CLEAN to a two-dimensional fluctuation power spectrum is shown in Fig.\ \ref{exclean}.  This spectrum was derived from VLA observations which will be discussed in  detail in the following section.  Fig.\ \ref{exclean} first shows the IRF computed from the positions of the elements of the array, in this case, VLA antennas.  This is computed over an area twice as wide as the spectrum itself to facilitate shifting of the IRF to be centered anywhere within the spectrum being CLEANed.  One can see the main lobe is significantly elongated in the east-west direction, resulting from the hybrid VLA configuration used (BnA; see Sec.\ 3.1).  One may also notice two relatively large sidelobes to the east and west of the main lobe, as well as many other significant sidelobes.\par
The set of panels below the IRF image show the CLEANing process.  Among these, the initial spectrum is shown in the upper left with the peak pixel marked with a white $\times$.  The panels directly to the right of this one show the spectrum after successive CLEAN iterations as significant components are convolved with the IRF and subtracted.  To make this illustration more compact, the CLEAN gain was set relatively high ($=$0.5), leading to convergence in only three iterations.  The final difference image is shown in the lower left panel with an image of the restored CLEAN components directly to its right.  The combination of these two images forms the final CLEANed spectrum, shown in the lower right panel.  One can see that the significant source apparent within the spectrum is no longer as elongated in the east-west direction due to the mitigation of the two main sidelobes.  In addition, the peak S/N was improved by a factor of 1.14, indicating a general reduction in the influence of sidelobes.\par
Following such an application of CLEAN, the resulting CLEAN components can be used to generate model correlations (i.e., $\widehat{C}$) used to solve for new values for the element-based gains.  A new image can then be made and the whole process is repeated until it converges (i.e., the image does not change appreciably from one iteration to the next).  When properly constrained (see above), this process has been shown to be quite robust and weakly dependent on the quality of the initial model \citep[again, see][]{cornwell99}.  The fidelity of this model only really impacts how quickly the process converges, but does not significantly affect the final result.

\section{Test Data}
\subsection{VLA Data}
To demonstrate the method described in Sec.\ 2, we have applied it to data from two different sparse arrays located in New Mexico.  The first is the VLA, an interferometric radio telescope consisting of 27 dish antennas, each 25 m in diameter, arranged in an inverted ``Y'' pattern (located at $34.079^\circ$N, $107.618^\circ$W).  The array is cycled through four configurations, A, B, C, and D spanning roughly 40, 11, 6, and 1 km, respectively.  Hybrid configurations are used for brief periods of time where the northern arm is set at a larger configuration than the southeastern and southwestern arms for observing southern sources that are always relatively close to the southern horizon.  The array is optimize for the microwave regime, but had, until 2008, two VHF systems at 330 and 74 MHz.  Upgraded, broader-band replacement systems are currently being commissioned.\par
It is these VHF systems that are extremely sensitive to ionospheric disturbances, even small-scale turbulent fluctuations.  Specifically, the phases of the visibilities measured for each baseline are proportional to the difference in TEC along the lines of sight of the two antennas, or $\delta\mbox{TEC}$.  When observing a bright source with either VHF system,   $\delta\mbox{TEC}$ can be measured to a precision of $<\!10^{-3}$ TECU \citep{helmboldt12a,helmboldt12c,helmboldt12d}.  For the 74 MHz system in particular, \citet{helmboldt12d} found a linear relationship between $\delta\mbox{TEC}$ precision and source peak intensity, where $\delta\mbox{TEC RMS}\simeq 0.09 I_p^{-1} \mbox{ TECU}$ with $I_p$ in units of Jy beam$^{-1}$ and $I_p<750$ Jy beam$^{-1}$.  For sources brighter than 750 Jy, the achieved precision is roughly constant at 1--2$\times10^{-4}$ TECU; a similar precision has been achieved for such exceptionally bright sources with the 330 MHz system \citep{helmboldt12a}.\par
Note that the sensitivity of an interferometer like the VLA to the TEC gradient rather than TEC makes it somewhat distinct from other similar remote sensing instruments.  Unlike other radio-frequency sensors that use a pulsed or otherwise coded signal to measure the delay affected by the ionosphere to estimate TEC (e.g., GPS), the cosmic radio sources observed by the VLA generally emit temporally uniform signals.  It is only by comparing the received signal at two different locations that one can detect the effect of ionospheric structure as these two copies of the cosmic signal will be more out of phase than they otherwise would be due to ionospheric structure.  If the ionosphere was a uniform slab of ions, this extra phase difference would not materialize.\par
{\bf 
It is this fact that only allows one to measure $\delta\mbox{TEC}$ with an interferometer like the VLA.  Even when one measures this quantity relative to a common reference antenna, it is difficult to perform a generalized spatial and temporal spectral analysis based on the $\delta\mbox{TEC}$ time series.  This is because TEC along the line of sight of the reference antenna also varies with time in a way that is comparable to, but slightly different from the antenna to which it is being compared.  Basically, the reference antenna TEC  does not constitute a stable reference TEC level.  In other words, $\delta\mbox{TEC}$ is \emph{not} analogous to relative TEC measured with GPS data that shows TEC variations relative to a fairly stable instrumental bias level.\par
This is why a $\delta\mbox{TEC}$ time series does not directly reflect variations in TEC, but rather changes in the TEC gradient.  That is to say, if there were temporal variations in the local TEC that affected all antennas in virtually the same way, all measured $\delta\mbox{TEC}$ values would be essentially zero and would show no evidence of fluctuations at all.  Thus, there must be a non-negligible spatial TEC gradient over the array at any given time for it to be affected and yield meaningful measurements of $\delta\mbox{TEC}$.\par
In the specific case of the VLA, a procedure to extract the full, two-dimensional TEC gradient at each antenna as a function of time from high-precision $\delta\mbox{TEC}$ measurements was developed by \citet{helmboldt12a}.  This procedure relies on a two-dimensional polynomial fit to the $\delta\mbox{TEC}$ values measured on all baselines as a function of relative antenna positions to compute the full TEC gradient at each time step.  The polynomial fit is used rather than numerical/finite differencing methods because of the inverted ``Y'' shape of the VLA.  This is discussed in much more detail by \citet{helmboldt12a} and \citet{helmboldt12b}.
}\par
\begin{figure}
\noindent\includegraphics[width=6in]{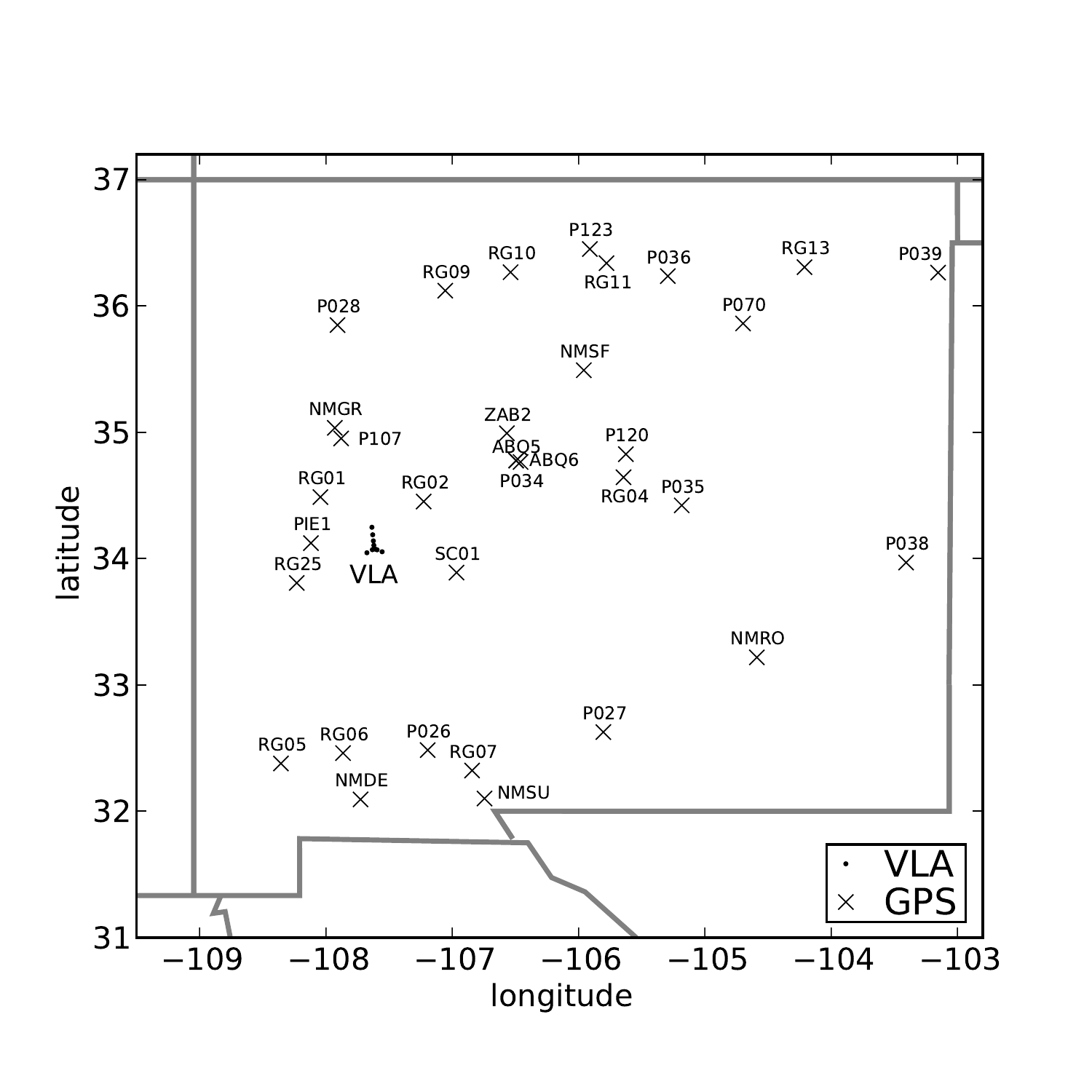}
\caption{The locations (geographic latitude and longitude) of the arrays used; the locations of the VLA antennas (BnA configuration; see Sec. 3.1) are shown as black points and the GPS receivers used are represented by black $\times$'s with their four-character station designations.  State/country boundaries are shown in grey.}
\label{arrays}
\end{figure}

The data used in this case study were retrieved from commissioning observations with the new broad-band P-band system, spanning 236--492 MHz.  The data were observed under VLA program ID TDEM0017 on 14 September 2012 starting at about 06:15 UT (23:04 local time) and are publicly available via the National Radio Astronomy Observatory (NRAO) data archive system (https://archive.nrao.edu/archive).  At this time, the VLA was in the hybrid BnA configuration, with the north arm set a the A configuration length (roughly 20 km) and the other two arms at the B configuration length (roughly 6 km).  The layout of the array along with the GPS receivers discussed in the following section are shown in Fig.\ \ref{arrays}.  At this time, P-band systems were only usable for 10 of the 27 VLA antennas.  The observations were focussed on two galaxy clusters, Abel 2256, or ``A2256,'' and CIZA J2242.8+5301, also known as the ``sausage'' cluster.  The data analyzed for this study used only one of two available sub-bands, spanning 236--364 MHz.  The clusters were observed in ten $\sim\!\!35$ minute segments, alternating between the two clusters from one segment to the next.  Measurements of $\delta\mbox{TEC}$ were obtained between each antenna and a reference antenna at intervals of 8 s, and were de-trended to remove instrumental effects by subtracting a linear fit within each $\sim\!\!35$-minute segment.  These were converted to measurements of the two components (north-south and east-west) of the TEC gradient at each antenna according to \citet{helmboldt12a}.\par
Because each of the two observed galaxy clusters provided significantly different lines of sight through the ionosphere, the spectral analysis described in Sec.\ 2 was performed separately for each $\sim\!\!35$-minute segment.  Within each segment, for each antenna and each component of the gradient, the time series was Fourier transformed to generate values of $\widetilde{T}(\nu)$ with a DFT using 30 frequencies up to 26.4 hr$^{-1}$.  The 8 s sampling interval allows for higher frequencies to be used, but we found no evidence of significant fluctuations for $\nu\!>\!26.4$ hr$^{-1}$.  The DFT was computed 37 times for each segment, using a sliding window that was $N-37$ samples wide, where $N$ is the total number of samples in the segment, which varied from 208 to 271.  Thus, for each segment, gradient component, and antenna, $\widetilde{T}(\nu)$ was measured at 8 s intervals, with the first and last intervals separated by roughly five minutes (i.e., $37\times8\mbox{ s}\!=\!4.93\mbox{ m}$).  These measurements were then used to compute the correlations among the 45 unique baselines within each segment and for each gradient component and frequency, $\nu$, by averaging the product $\widetilde{T}_j(\nu)\widetilde{T}_k^\ast(\nu)$ over this $\sim$5-minute span.

\subsection{GPS Data}
During the VLA observations described in Sec.\ 2.1, there were 33 continuously operating GPS receivers within New Mexico, shown on the map in Fig.\ \ref{arrays} with their four-character station designations.  Data from these receivers is publicly available and was obtained in compressed RINEX format from the NOAA (ftp://www.ngs.noaa.gov/cors/rinex/) and CORS (ftp://data-out.unavco.org/pub/rinex/) databases.  Values for relative slant TEC (STEC) were extracted from these data using the GPS Toolkit \citep[GPSTk;][]{munton04} using both the pseudorange and smoothed range values.  However, after performing the full analysis using both these values, there was no apparent difference in the results, and we consequently only used the pseudorange-based STEC measurements.  GPSTk was also used to compute the latitude and longitude of the ionospheric pierce point (IPP) for each satellite/receiver pair at each time, as well as a slant-to-vertical TEC correction factor, both assuming an altitude of 300 km.\par
The STEC values were de-trended according to \citet{hernandez06} and \citet{helmboldt12b} by subtracting from each time step the mean STEC between the time step 600 s before it, and the one 600 s after it.  Those time steps less than 600 s from the beginning or end of any contiguous segment for a particular satellite/receiver pair were therefore excluded.  This de-trending scheme optimizes the response of the resulting temporal spectra for periods $>\!10$ minutes, which covers the range spanned by medium and large-scale TIDs \citep{hernandez06,shiokawa08,borries09,helmboldt12b,helmboldt12d}.\par
For consistency with the VLA observations, the de-trended GPS STEC data were processes within the same $\sim\!\!35$-minute segments.  Within each segment, the time series for every satellite/receiver pair that had de-trended STEC values for at least half of the time was Fourier transformed with a DFT for 12 temporal frequencies up to 6 hr$^{-1}$.  This was done using a sliding window that was $\Delta\mbox{t}-5\mbox{ minutes}$ wide, where $\Delta\mbox{t}$ was the temporal width of the segment, and was stepped 30 s at a time.  This produced 10 values of $\widetilde{T}(\nu)$ for each segment, satellite/receiver pair, and frequency over a five-minute period so that a similar integration time (5 m) would be used to generated the correlations for the GPS data as was used for the VLA data.\par
Following this, correlations were computed two ways.  First, within a single segment, correlations among all receivers were computed separately for each satellite that was visible by at least 20 receivers during the segment (usually, about 10 satellites per segment).  For these, no slant-to-vertical correction was applied because this correction varies insignificantly over the 33-element GPS array for a single satellite.  Next, within each segment, correlations were generated among all available satellite/receiver pairs, between 272 and 351 pairs per segment.  For these, since the lines of sight spanned a wide range of elevations, the mean slant-to-vertical correction per satellite was applied before correlating.  While this does not take into account the effect wavefront orientation will have on the measured TEC values over such a large area, it does at least provide a good approximate geometric correction.

\section{Spectral Analysis}
\subsection{Examples of the Method}
For the VLA and GPS-based correlations described in Sec.\ 3.1 and 3.2, we developed python-based software, relying heavily on the NumPy package (http://www.numpy.org), to apply the spectral analysis detailed in Sec.\ 2.  Specifically, for each array, time segment, and temporal frequency, the correlations were binned in a uniform, two-dimensional grid in the $\alpha,\beta$ plane and, following equation (\ref{final}), the power spectrum was computed using a two-dimensional fast Fourier transform (FFT) of the gridded correlations.  An FFT-based approach was chosen because within the iterative deconvolution process, the spectral map for each temporal frequency has to be generated several times, and the FFT helps streamline this process considerably.  By summing correlations within the $\alpha,\beta$ grid, the FFT provides a method for performing a nearly equivalent sum to the one in equation (\ref{final}) in a fraction of the time.\par
\begin{figure}
\noindent\includegraphics[width=6in]{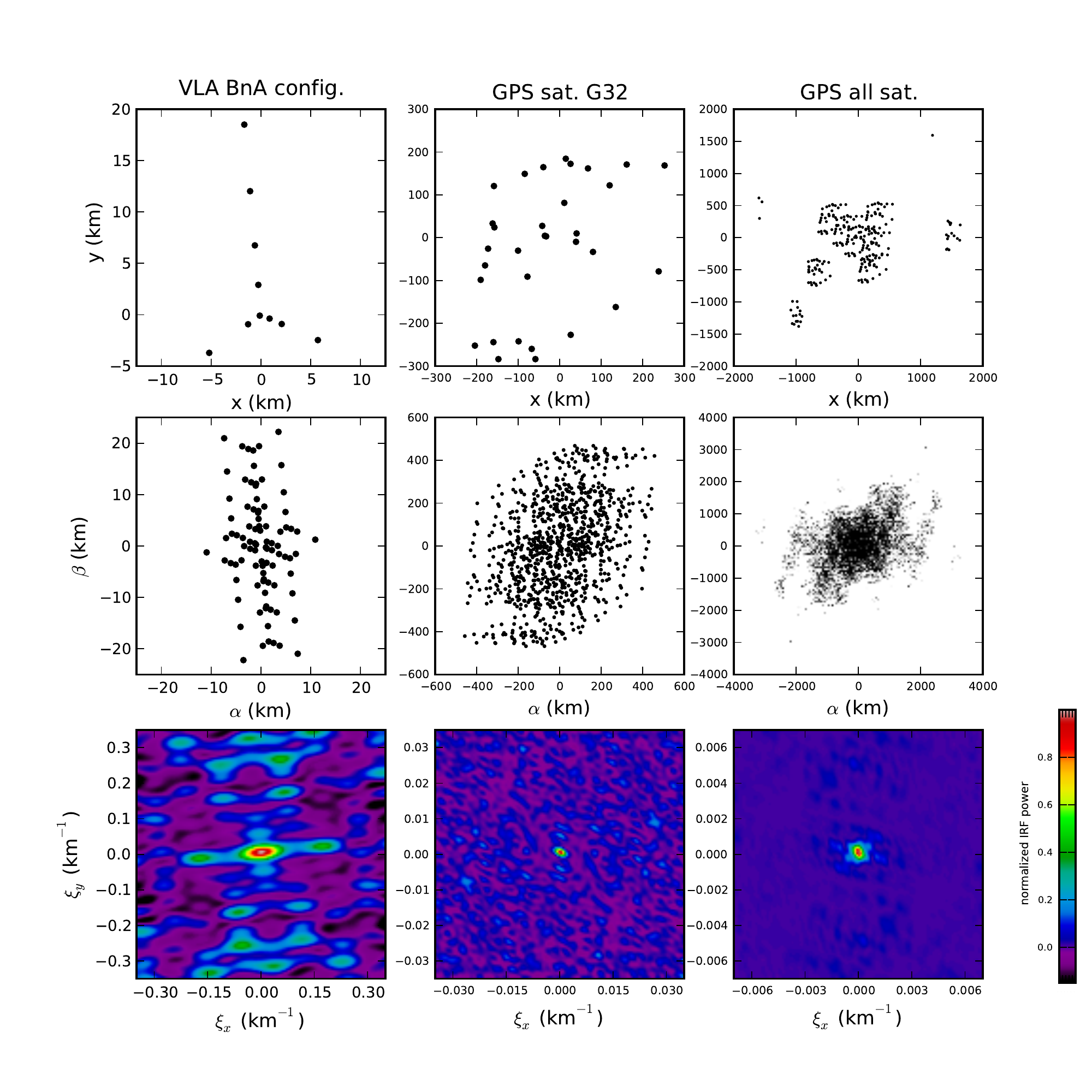}
\caption{Examples of (top) array layouts, (middle) resulting $\alpha,\beta$ plane coverage, and (bottom) impulse response functions for (left) a VLA observation of a single source, (middle) a GPS observation of a single satellite, and (right) a GPS observation of all available satellites for a single $\sim\!\!35$-minute segment.}
\label{psf}
\end{figure}

The $\alpha,\beta$ grid was set up to be coarse enough to produce spectral maps that grossly oversampled the power spectra, i.e., $\delta\alpha = (N_p \delta\xi_x)^{-1}$ [and $\delta\beta = (N_p \delta\xi_y)^{-1}$], where $N_p$ is the number of pixels along one side of the spectral map and $\delta$ indicates the width of a grid cell in either the $\alpha,\beta$ or $\xi_x , \xi_y$ plane.  This was intended to facilitate the CLEANing process, that is, to ensure that any peaks within the spectral maps would lie as close as possible to the center of a pixel.  For each array, 512 pixels were used with maximum spatial frequencies of 0.35, 0.035, and 0.007 km$^{-1}$ for the VLA BnA configuration, the single-satellite GPS data, and the multi-satellite GPS data, respectively.  For our implementation of the CLEAN algorithm, we computed the nominal IRF for each array on a $1024\times1024$ grid with the same pixel scale as the actual spectral maps so that the IRF could be shifted to any location and still cover then entire spectral map.  Fig.\ \ref{psf} shows the array layouts, $\alpha,\beta$ plane coverages, and resulting IRFs for the VLA, a single GPS satellite (G32 at about 09:50 UT), and all visible satellites (again, around 09:50 UT).\par
\begin{figure}
\noindent\includegraphics[width=4in]{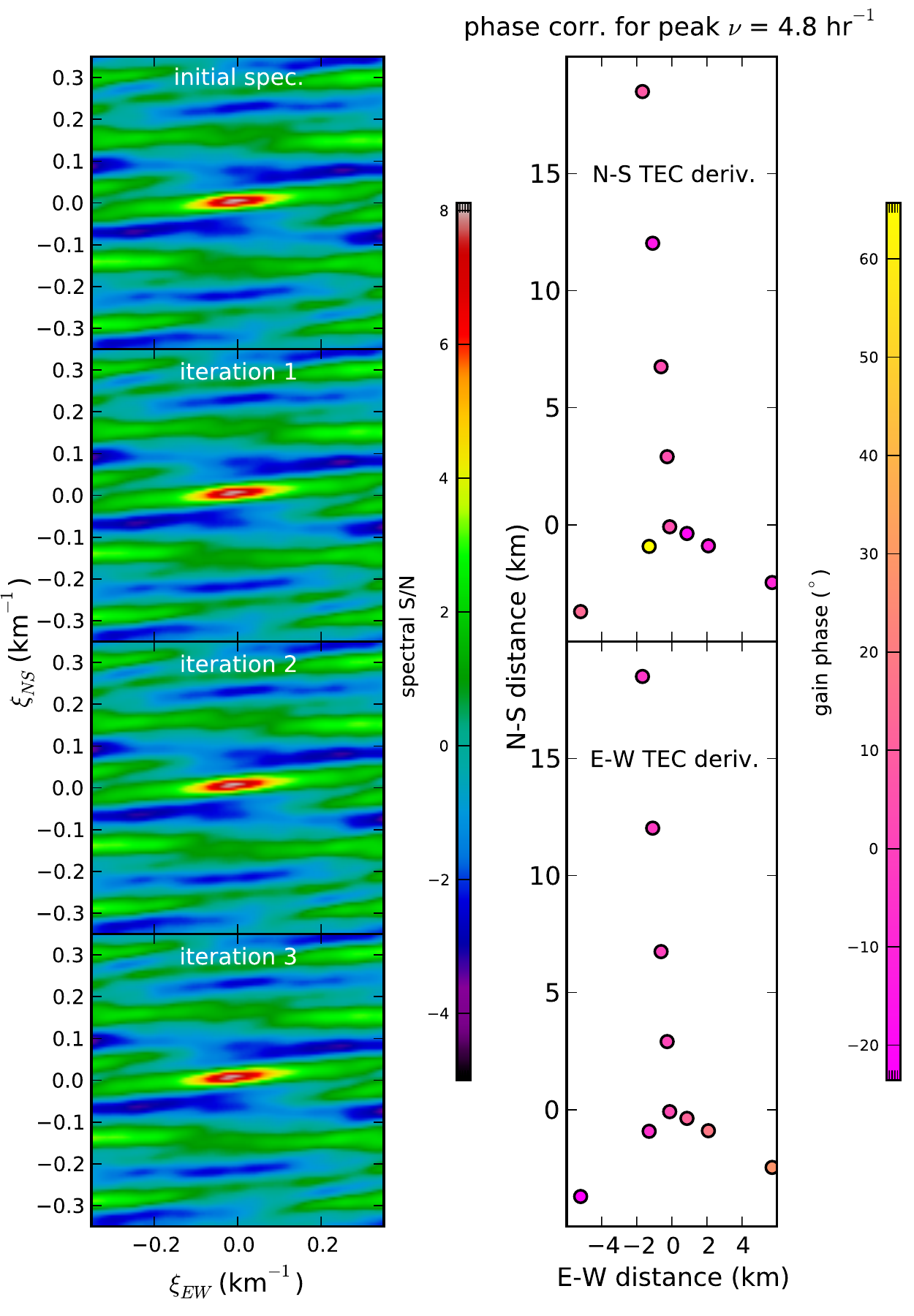}
\caption{The steps in the deconvolution process for one $\sim\!\!35$-minute segment of VLA observations:  (left) the mean TEC gradient spectral power over all temporal frequencies at each iteration in the deconvolution process, normalized by the median absolute deviation (MAD); (right) the final phase correction for each VLA antenna for the peak temporal frequency (4.8 hr$^{-1}$) for each component of the TEC gradient (i.e., north-south, and east-west partial derivatives).}
\label{sausage}
\end{figure}
\begin{figure}
\noindent\includegraphics[width=6in]{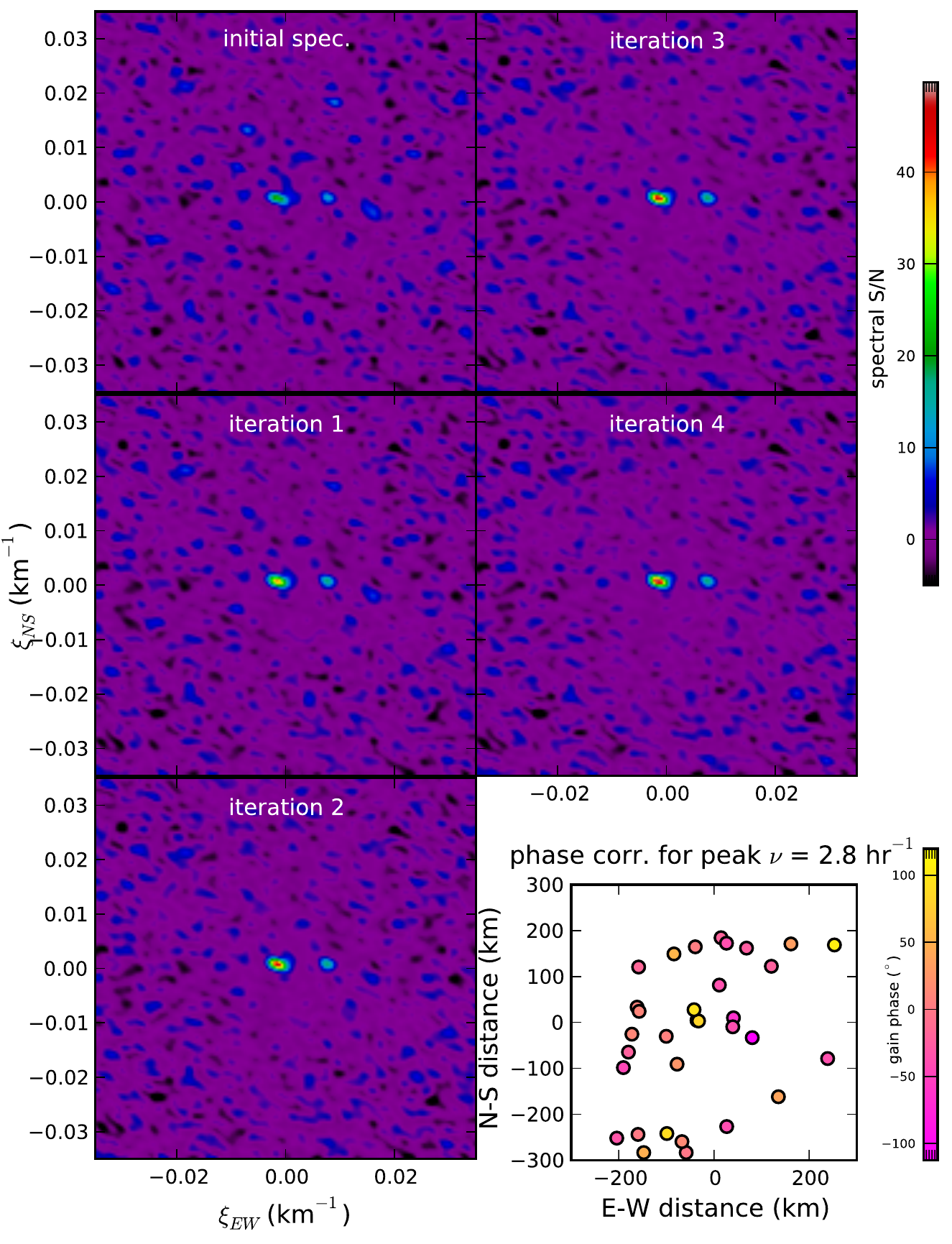}
\caption{Similar to Fig.\ \ref{sausage}, the deconvolution process as applied to GPS data for a single satellite (G32) during the same $\sim\!\!35$-minute period as Fig.\ \ref{sausage}.  The final phase corrections are shown for the peak frequency, 2.8 hr$^{-1}$.}
\label{g32}
\end{figure}
\begin{figure}
\noindent\includegraphics[width=3.5in]{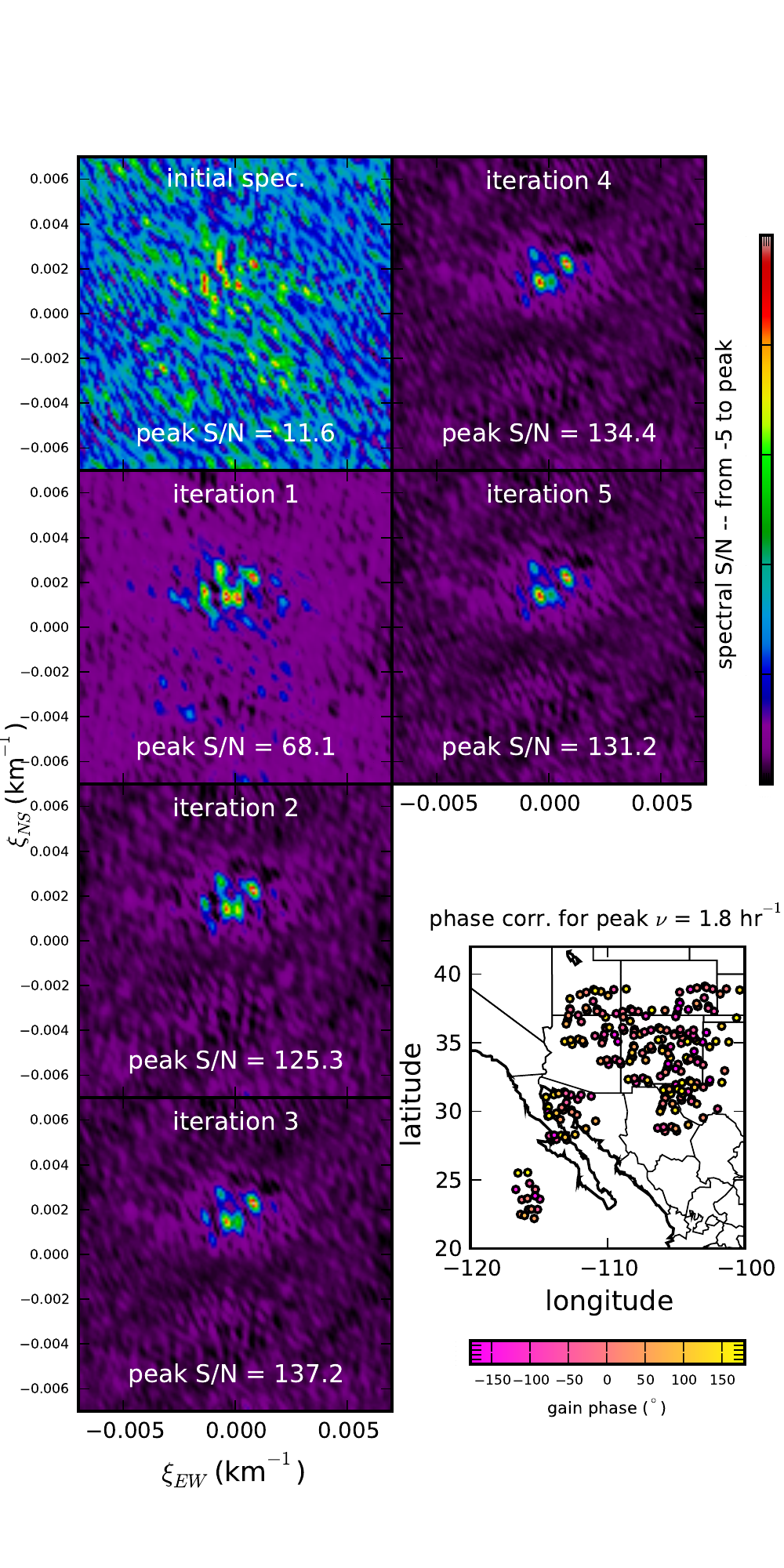}
\caption{The same as Fig.\ \ref{g32}, but for the combined analysis of all satellite/receiver pairs.  For the spectral images, because of the dramatic change in S/N during the deconvolution process, each image is normalized by its own maximum S/N, which is printed in each panel; the minimum for all the images is set to $-5$.}
\label{t07}
\end{figure}

Our implementation of the self-calibration algorithm was patterned after that used within the DIFMAP \citep{shepherd95} software package for synthesis imaging (ftp://ftp.astro.caltech.edu/pub/difmap/difmap.html), which uses a gradient search to determine the complex gains that minimize $\Delta$ as defined in equation (\ref{fit}).  As explained in Sec.\ 2.2, we used phase-only self-calibration throughout this analysis to constrain the fit, keeping it from fitting the noise.  Examples of the application of this algorithm in concert with iterations of imaging and CLEANing are shown in Fig.\ \ref{sausage}--\ref{t07}.  In each figure, the initial CLEANed map is shown, followed by CLEANed maps made after the application of gain corrections derived from successive iterations of self-calibration.  In each case, the parameters of the Gaussian IRF used to restore the CLEAN components was determined by fitting a two-dimensional Gaussian to the main lobe of the IRF.  The parameters of these Gaussian IRFs for the VLA, single-satellite GPS, and multi-satellite GPS, respectively, were major axes of 0.11, 0.0030, 0.0007 km$^{-1}$, minor axes of 0.034, 0.0015, 0.0005 km$^{-1}$ and position angles (clockwise from north) of $82^\circ$, $120^\circ$, $150^\circ$, where the axes widths are the full width at half maximum (FWHM).\par
In all cases, each temporal frequency was CLEANed and calibrated separately, but the maps in Fig.\ \ref{sausage}--\ref{t07} show the average power over all frequencies.  For the VLA maps, both components of the TEC gradient were analyzed separately and added together such that a wave with a TEC amplitude $A$ and wavelength $\lambda$ would have a peak spectral power of $(2\pi A / \lambda)^2$ in the resulting map.  All maps were normalized by the peak of the corresponding IRF, which is denoted in the reported spectral power units as IRF$^{-1}$.  In each figure, the final phase corrections from the element-based gains are displayed for the peak temporal frequency as color-coded plots of the element positions.\par
For the VLA spectral maps, one can see that the phase-only self-calibration process converges rather quickly, with no appreciable difference between the initial spectrum and the one after the second round of phase-only self-calibration.  Except for one isolated antenna, the phase components of the element-based gains seem to vary little among the array elements in this case.\par
The single-satellite example shown in Fig.\ \ref{g32} shows that, first, more iterations are typically need for phase-only self-calibration to converge for the GPS array.  This is not particularly surprising given that it spans a substantially larger area and is therefore more prone to be affected by localized wavefront distortions.  Because of its larger size and because the observed sources (GPS satellites) move much faster on the sky than cosmic sources, the GPS array is also more prone to errors in the de-trending process that is intended to account for the effects of the changing line of sight of each receiver toward a particular satellite through the ionosphere.  This leads to what are effectively gain errors in the computed complex quantities $T(\nu)$, which are mitigated by the self-calibration process (see Sec.\ 5 for more discussion).\par
One can see in Fig.\ \ref{g32} that, as expected, the peak S/N gradually improves with each successive iteration.  One can also see several moderately significant peaks (S/N$\sim$10) in the initial spectrum that turn out to be purely the result of phase distortions/errors as they substantially weaken/disappear with sequential applications of self-calibration.  In the end, the peak spectral S/N was improved by a factor of $\sim\!\!2$.  The phases of the final receiver-based complex gains for the peak temporal frequency show now particular pattern, implying some combination of measurement/de-trending errors and relatively small-scale wavefront distortions.\par
The multi-satellite data provide the most dramatic illustration of the utility of this spectral analysis to detect and characterize wave signatures in the presence of significant distortions and/or measurement errors.  As Fig.\ \ref{t07} shows, the improvement in S/N over the entire deconvolution process is so dramatic that the color map had to be redefined for each panel.  One can see in the initial spectrum that several significant and many marginal detections of wave signatures are apparent.  With a single application of phase-only self-calibration, many of the marginal detections disappear and the peak S/N increases by more than a factor of six.  Subsequent iterations of phase-only self-calibration reveal all of these structures to be the result of phase errors/distortions, converging on a group of roughly northward propagating, relatively large waves (wavelengths $\sim\!\!400$--$1,000$ km).  The final peak S/N is more than a factor of 10 better than in the initial spectral map.  Similar to the single-satellite result, the final phase corrections show no clear pattern over the array.
\subsection{Spectral Maps}
\begin{figure}
\noindent\includegraphics[width=6in]{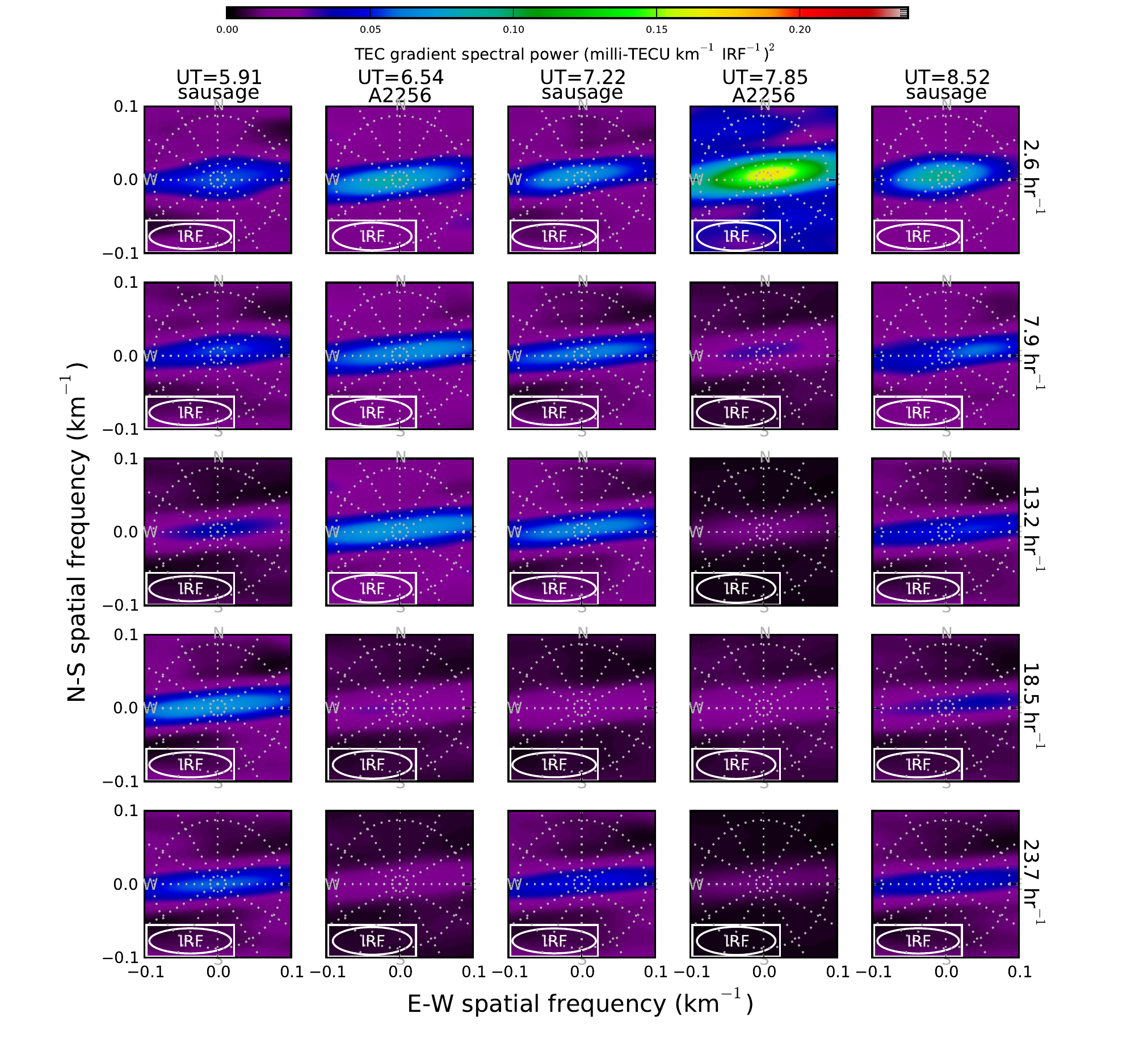}
\caption{Images of the mean, combined spectral power of the TEC gradient from the VLA data within bins of temporal frequency (see labels on the right) for the first five $\sim\!\!35$-minute segments.  {\bf Each power spectrum is the sum of the spectra measured separately for the two components of the TEC gradient such that a wave with TEC amplitude $A$ and wavelength $\lambda$ with have a peak power of $(2\pi A / \lambda)^2$ (see Sec. 4.1).} The name of the observed cosmic source is printed above each column.  A polar grid is drawn for reference with spatial frequency increasing (wavelength decreasing) with radius.  The oblong shape of the IRF for the VLA BnA configuration (see Sec.\ 3.1) is illustrated with ellipses with dimensions equal to the full width at half maximum (FWHM) of the IRF.}
\label{chmap1}
\end{figure}

\begin{figure}
\noindent\includegraphics[width=6in]{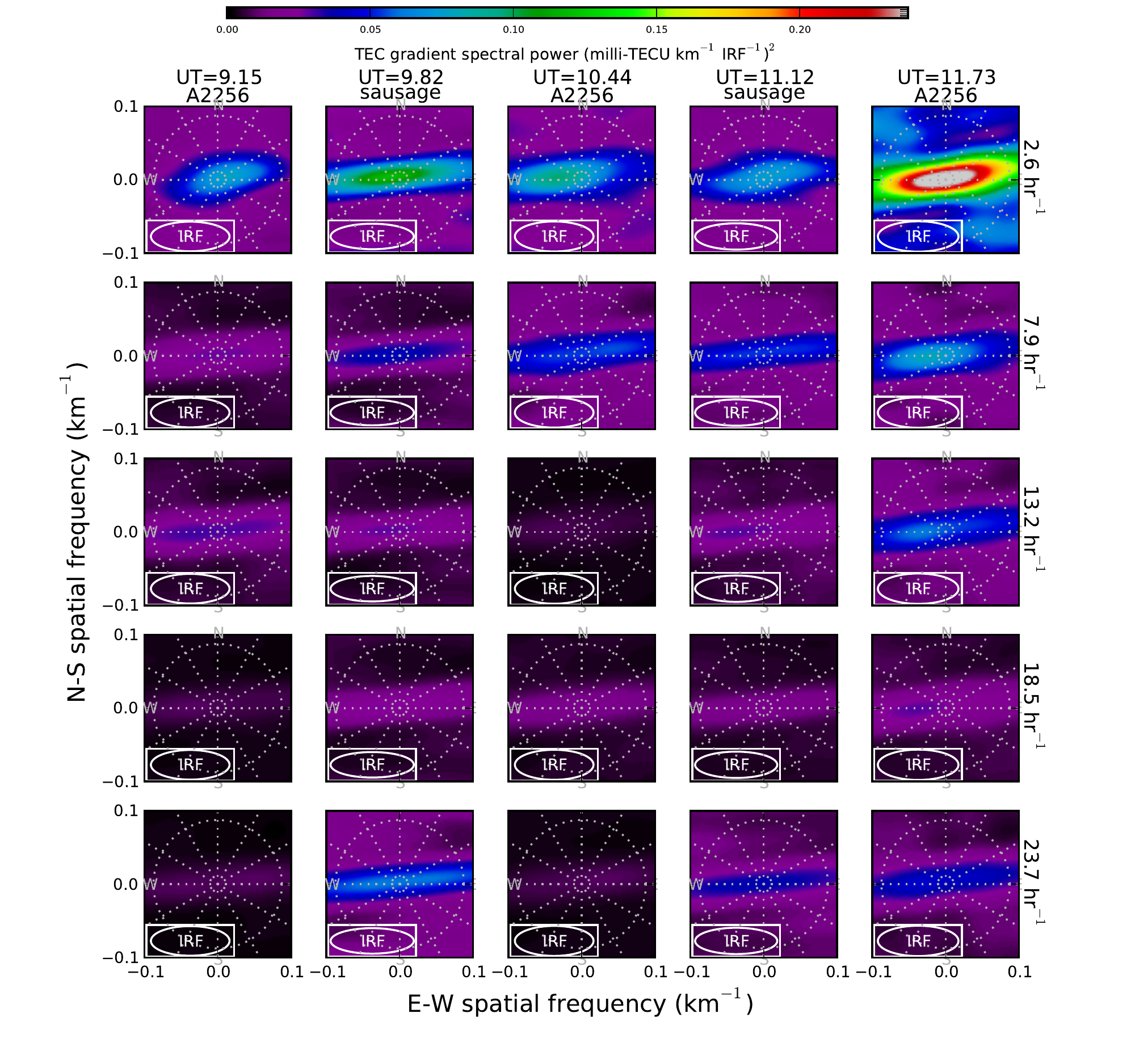}
\caption{The same as Fig.\ \ref{chmap2}, but for the last five $\sim\!\!35$-minute segments.}
\label{chmap2}
\end{figure}

\begin{figure}
\noindent\includegraphics[width=6in]{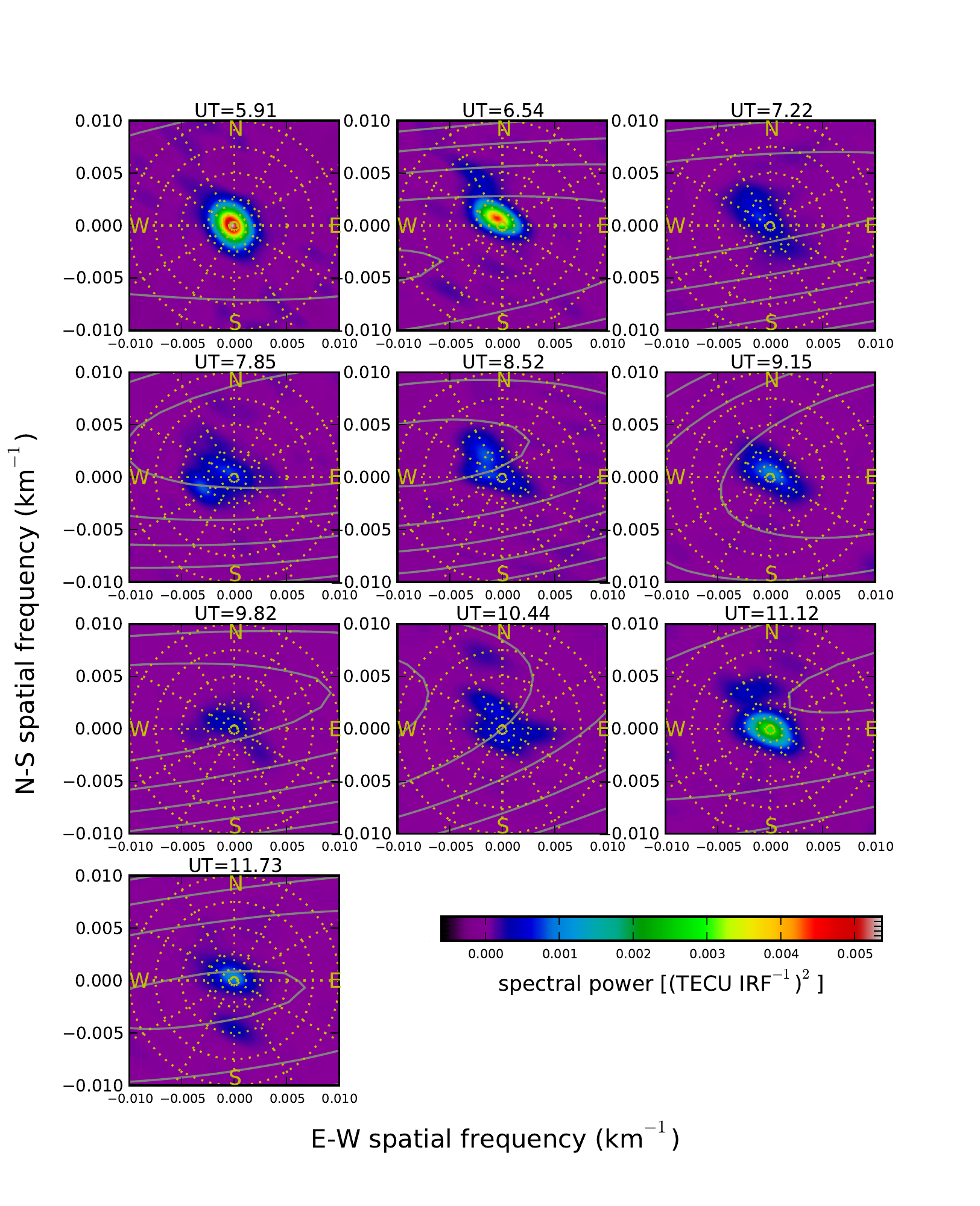}
\caption{Images of the mean spectral power over all temporal frequencies from analysis of observations of individual satellites, averaged over all satellites (typically $\sim\!\!10$ per image) within each $\sim\!\!35$-minute interval.  The contours show the mean combined spectral power of the TEC gradient from VLA data within the same time intervals.  As in Fig.\ \ref{chmap1} and \ref{chmap2}, a white polar grid is given for reference.}
\label{single}
\end{figure}

The final CLEANed and self-calibrated spectral maps are show in Fig.\ \ref{chmap1}--\ref{all}.  The average spectral maps within bins of temporal frequency are shown for the VLA data separately for each $\sim\!\!35$-minute segment in Fig.\ \ref{chmap1} and \ref{chmap2}.  The bins are roughly 5.3 hr$^{-1}$ wide, with the first bin covering nearly all of the range in temporal frequency covered by the GPS-based spectra.  The observed cosmic source (sausage cluster or A2256) is printed above each column.  In each panel, a polar grid is printed for reference and the approximate size and shape of the CLEAN IRF is shown with an ellipse in the lower left.  In the lowest temporal frequency bin, there are detections of waves throughout the observing run.  These vary in strength with peaks at 07:51 and 11:44 UT (00:41 and 04:33 local time).  While the oblong nature of the IRF complicates matters somewhat, the strongest of these waves appear to be moving roughly to the northeast while some of the weaker waves seem to be directed westward.  Higher temporal frequency wave are also detected intermittently, at times rivaling the lower-frequency waves in spectral power.  These seem to be mostly directed either eastward or westward.\par
The mean spectral maps over all frequencies and satellites for the single-satellite GPS data are shown in Fig.\ \ref{single}.  Also plotted are contours for the mean spectral power from the VLA data for $\nu\!<\!6$ hr$^{-1}$, the maximum temporal frequency used for the GPS data.  In some cases, the directions of the small-scale waves detected with the VLA roughly match those of the larger waves detected with the GPS array (see UT$=$8.52, 10.44), implying that they may represent smaller-scale structures associated with the larger waves.  However, in most cases, the waves detected with the VLA seem to be moving in unrelated directions, highlighting the usefulness of joint observations with a relatively large GPS-based array and a small-scale, high-precision array like the VLA.  In addition to superior TEC fluctuation sensitivity, the fact that the VLA measures the TEC gradient rather than actual TEC biases it toward smaller-scale waves because their amplitudes are effectively multiplied by $2\pi\xi$, making it able to sense smaller disturbance beyond the capabilities of GPS-based arrays.  Conversely, the shear size of the GPS array renders it capable of detecting large-scale waves (i.e., small $\xi$) to which the VLA is simply not sensitive.\par
The single and multi-satellite GPS-based spectral maps are compared in Fig.\ \ref{all} where the mean spectral power over all $\nu$ for each time segment is displayed as an image for the multi-satellite data and as contours for the single-satellite data.  For the single-satellite data, the spectral maps are again averaged among all satellites as in Fig.\ \ref{single}.  One can see that, as may be expected, the main advantage of analyzing all satellite/receiver pairs together is improved spectral resolution.  For the most part, the distribution of spectral power in the multi-satellite maps agrees with that of the single-satellite maps, but with much better detail.  In many cases, what appears to be one or two significant wave features in the single-satellite map is revealed to be a series of several waves traveling in significantly different directions.\par
There is a notable exception to the general agreement between single and multi-satellite spectra.  This exception involves the time segment centered on 07:51 UT, which shows a strong detection of a southwestward-directed, $\sim\!\!300$ km wave within the mean single-satellite spectrum that has no counterpart in the multi-satellite spectrum.  Upon further inspection, we found that this wave was only detected toward satellite G30, which was visible at the time nearly due south of the array at a low elevation of about $13^\circ$.  This wave peaks at a temporal frequency of 2.75 hr$^{-1}$ as apposed to the strong westward-propagating waves seen in the multi-satellite spectral map which peak at $\nu\!=\!1.75$ hr$^{-1}$.  After applying the final multi-satellite phase corrections to only the data for satellite G30 and making a new spectrum, the signature of the southwestward-propagating wave was essentially eliminated.  This implies that the self-calibration process effectively suppressed the signature of this wave because it was only detected within a relatively small portion of the full, multi-satellite array.  Thus, one should note that there is significant value in analyzing observations of individual satellites separately for the purpose of characterized localized, wavelike disturbances of limited spatial extent.

\subsection{Waveform Reconstruction}
The nature of the deconvolution technique described in Sec.\ 2.2 provides a path toward reconstructing the observed two-dimensional TEC or TEC gradient waveform.  This has scientific and practical value as it allows one to visualize what the disturbances look like apart from the spectra shown in Fig.\ \ref{sausage}--\ref{all}, and delivers a time-dependent map of TEC fluctuations that may be used, for example, to correct for ionospheric distortions within wide-field astronomical images.  Note, however, that line-of-sight effects may limit the effectiveness of such an application to relatively large fields of view where the lines of sight toward different cosmic sources within such a field can be far from parallel.\par
The heart of this reconstructive capability is the CLEAN component model that is constructed as an essential part of the deconvolution process.  Each CLEAN component represents a single spatial Fourier mode that, along with its corresponding temporal frequency, can be used to evolve it and all other components over a given spatial extent and time period.  The only necessary parameters for the CLEAN components that are not yielded by the deconvolution process are the phase offsets of their Fourier modes relative to one another.  Fortunately, because we have access to the original $\widetilde{T}(\nu)$ values (see Sec.\ 2 and 3), we can determine these phase offsets through a simple linear fit.  Specifically, for array element $j$, we assumed
%\begin{linenomath}
\begin{equation}
\widetilde{T}_j(\nu) = g_j(\nu) \sum_{k=1}^{N_{cc}} A_k e^{2\pi i(\xi_{x,k} x_j + \xi_{y,k} y_j)}
\label{recon}
\end{equation}
%\end{linenomath}
where $N_{cc}$ is the number of CLEAN components for frequency $\nu$ and $A_k$ are a set of complex coefficients solved for with a standard linear least-squares fit using all array elements.  In practice, the amplitudes of the $A_k$ coefficients are set by the CLEAN component spectral powers, and the linear fit is used on only solve for their phases.  The combined complex values for these coefficients then allow one to construct a waveform upon a chosen temporal and spatial grid according to
%\begin{linenomath}
\begin{equation}
W(t,x,y) = \Re \left \{ \sum_{j=1}^{N_{\nu}} \sum_{k=1}^{N_{cc,j}} A_{k,j} e^{2\pi i (\xi_{x,k,j} x + \xi_{y,k,j} y - \nu_j t)} \right \}
\label{wave}
\end{equation}
%\end{linenomath}
\par
A reconstruction of both components of the TEC gradient derived in this way from VLA data during the last $\sim\!\!35$-minute segment (centered at 11:44 UT) is shown in a movie, available electronically.  The first frame of the movie is shown in Fig.\ \ref{vlampg}.  The gradient is shown over the full P-band field of view, which has a diameter of about $3.4^\circ$, or about 17.6 km at an altitude of 300 km.  This is comparable to the area spanned by the BnA configuration VLA.  The large, westward-propagating structures implied by the spectral maps shown in Fig.\ \ref{chmap2} are apparent.  But, there are also significant distortions on scales of one to a few kilometers that are not obvious from visual inspection of the spectral maps.\par
We have also applied this procedure to the multi-satellite GPS spectral data to reconstruct a waveform for the TEC fluctuations.  Unlike the VLA data, which alternates between two cosmic sources probing two different parts of the ionosphere, the GPS data essentially covered the same area throughout the observing period.  Thus, we can make a reconstruction for the entire observing run with these data.  However, because each $\sim\!\!35$-minute segment was analyzed separately, there is no guarantee that the resulting wave form will be continuous across segment boundaries.  To enforce smooth transitions throughout the reconstructed waveform, we repeated our spectral analysis for 9 additional segments with upper and lower boundaries centered on each of the main segments, effectively applying the analysis with a sliding window stepped by $\sim\!\!35/2$ minutes.  We then computed a separate waveform for all 19 segments and added them together using a triangular weighting function that peaked at the center of each segment and went to zero at the edges.\par
The resulting $\sim\!\!6$-hour waveform can be view as a movie, available electronically.  The first frame of the movie is show as Fig.\ \ref{gpsmpg}.  The reconstruction was performed for the approximate field of regard of the GPS array assuming a minimum satellite elevation of $10^\circ$ and an ionospheric altitude of 300 km.  As the spectral analysis illustrated in Fig.\ \ref{single} and \ref{all} implies, there was a rich variety of wave-like fluctuations that was constantly evolving during this time period.  While the fluctuations are as large as $\sim\!\!0.2$ TECU, the typical amplitude is closer to 0.05--0.1 TECU, substantially smaller than would be noticeable on typical vertical TEC maps produced from GPS data that usually have uncertainties $\sim\!\!1$ TECU \citep[e.g.,][]{feldt11}.  There are some fluctuations in the periphery of the field of regard that are significantly larger, but are likely artifacts due to the poor satellite coverage in these parts of the map.

\subsection{Individual Waves}
The methods demonstrated above provide a useful basis for characterizing and visualizing ionospheric fluctuations sensed with sparse arrays.  However, the fluctuation spectral cubes generated constitute a relatively large amount of information to be searched, especially for the purposes of identifying individual waves and analyzing their properties.  Thus, it is impractical to do this ``by hand'' and an automated approach is preferred (or required for a big database of observations).  In this section, we demonstrate one possible method for such an automated analysis of the final CLEANed spectral cubes that uses existing, python-based software that is easily integrated with our own.  This analysis is essentially the same as that used in radio astronomy to identify/characterize individual, marginally resolved sources within CLEANed images.\par
While a CLEAN component model yielded by the deconvolution process provides a solid basis for reconstructing the observed field of TEC and/or TEC gradient fluctuations, it does not offer a suitable strategy for characterizing individual waves.  While they do not contribute significantly to the waveform reconstructions detailed in Sec.\ 4.3, relatively weak CLEAN components within the periphery of the spectral maps that are not associated with real waves can give the false impression of a population of small-scale, small amplitude waves.  In addition, a single wave may require multiple, closely spaced CLEAN components if phase distortions are not completely removed by the deconvolution process (see, e.g., Fig.\ \ref{exclean}).  A more prudent approach is to employ an algorithm that isolates ``islands'' of pixels of a minimum size and peak power, and then fits a two dimensional Gaussian to each, thus representing the image, or in this case, spectral map, as a sum of Gaussian components rather than delta functions.  One may isolate valid detections by limiting the peak power and size of the final fitted Gaussian using the spectral map RMS and IRF dimensions.\par
Most synthesis imaging software packages contain a routine for doing this analysis.  We have chosen to employ the Obit package \citep[][http://www.cv.nrao.edu/$\!\!\sim\!\!$bcotton/Obit.html]{cotton08}.  While Obit is designed to work with the commonly used Astronomical Image Processing System \citep[AIPS;][]{bride94}, it also has several independent routines that can read and write images and data in the Flexible Image Transport System \citep[FITS;][]{wells81} format and is fully scriptable in python, making it compatible with our python-based analysis software.  Specifically, the Obit task \texttt{FndSou} was used for each time segment and temporal frequency to identify and fit Gaussians to all islands with peak power $>\!5\sigma$, where $\sigma$ was the map RMS, and with a diameter larger than the FWHM of the Gaussian IRF minor axis.  For the VLA data, each component of the TEC gradient was analyzed separately.\par
\begin{figure}
\noindent\includegraphics[width=4in]{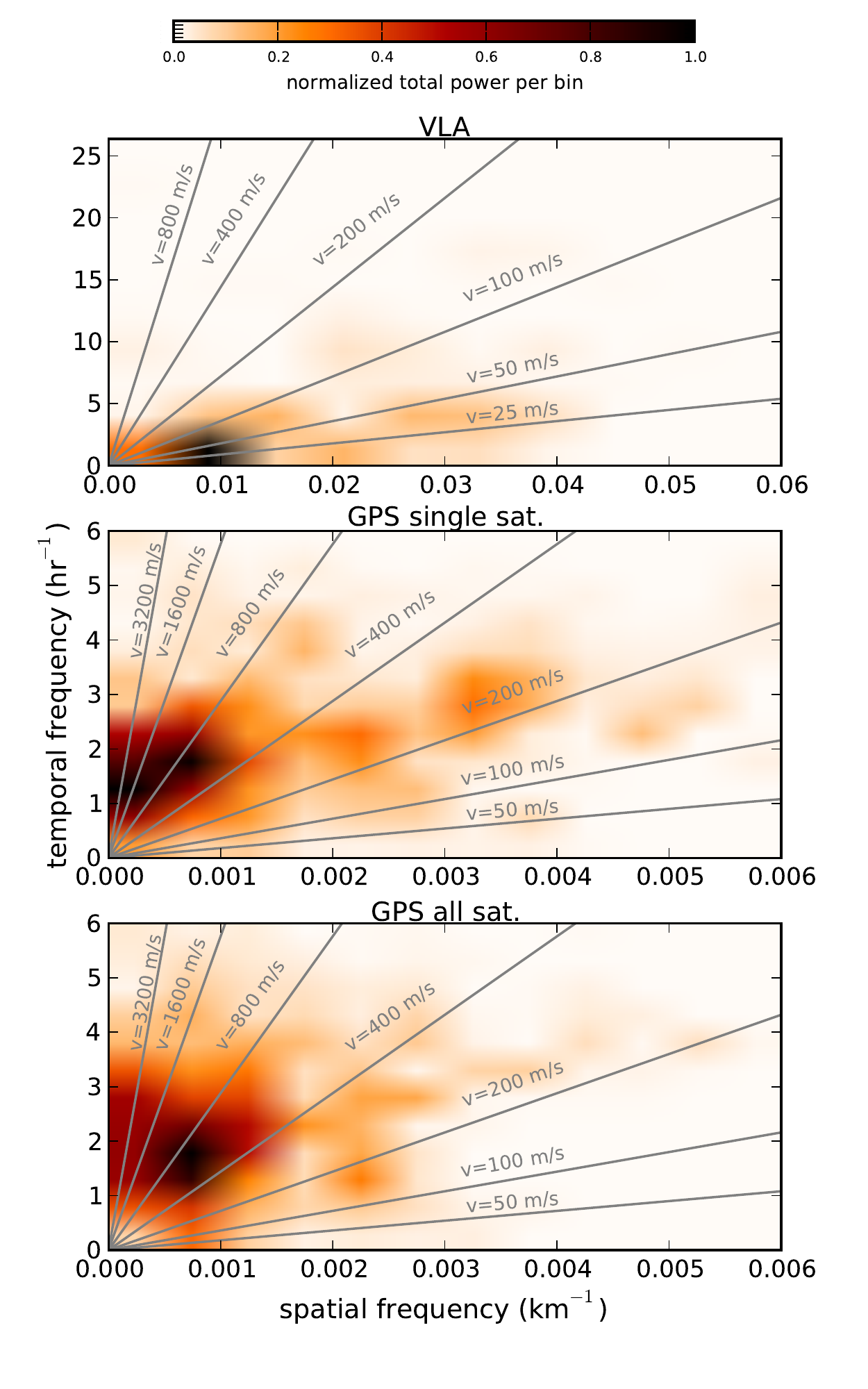}
\caption{Two-dimensional distributions (total power per bin) for the spatial (abscissa) and temporal (ordinate) frequencies of all detected waves found using the Gaussian fitting process described in Sec.\ 4.4.  Results are shown separately for (top) the VLA, (middle) GPS single-satellite, and (bottom) GPS multi-satellite data.  In each panel, lines of constant speed are drawn in grey.}
\label{waves}
\end{figure}

The two-dimensional distributions of temporal and spatial frequency of the waves detected with the Gaussian-fitting method are displayed in Fig.\ \ref{waves}.  These are shown separately for the VLA data, the single-satellite GPS data, and the multi-satellite GPS data.  In each panel, lines of constant speed (i.e., $\nu = v \xi$) are plotted.  From these one can see that the VLA data are dominated by a population of medium-scale ($\sim\!\!50$--100 km wavelength), relatively slowly moving ($\sim\!\!25$--50 m s$^{-1}$) waves.  However, there are also detections of smaller waves, about 20--30 km in wavelength traveling with roughly the same speeds.  In addition, there are higher-frequency waves, with wavelengths $\sim\!40$ km, moving faster at $\sim\!100$ m s$^{-1}$.  Both the single and multi-satellite GPS data show evidence of large-scale waves (wavelengths $\gtrsim\!\!1,000$ km) moving at speeds of about 800 to a few thousand m s$^{-1}$.  There is also a population of less prominent, but clearly detected medium-scale waves with wavelengths of about 300--500 km, moving at roughly 250 m s$^{-1}$ that are more obvious within the single-satellite data.  The fact that these waves are only marginally detected within the multi-satellite data likely attests to their limited physical extent as they were probably only detected toward one or a few satellites and were subsequently suppressed during the deconvolution of the multi-satellite spectra (see the discussion at the end of Sec.\ 4.2).

\section{Discussion and Conclusions}
We have detailed the successful adaptation of imaging and deconvolution techniques developed for radio astronomy to the spectral analysis of wavelike disturbances within Earth's ionosphere.  Using these techniques to correct for instrumental and physical phase distortions, we have demonstrated that even waves that are only marginally detected initially can be characterized with high precision by essentially focussing their spectral images.  We have also shown that this can be successfully achieved when using a relatively small, sparse array or for a larger array made up of several independently operating elements.\par
\begin{figure}
\noindent\includegraphics[width=6in]{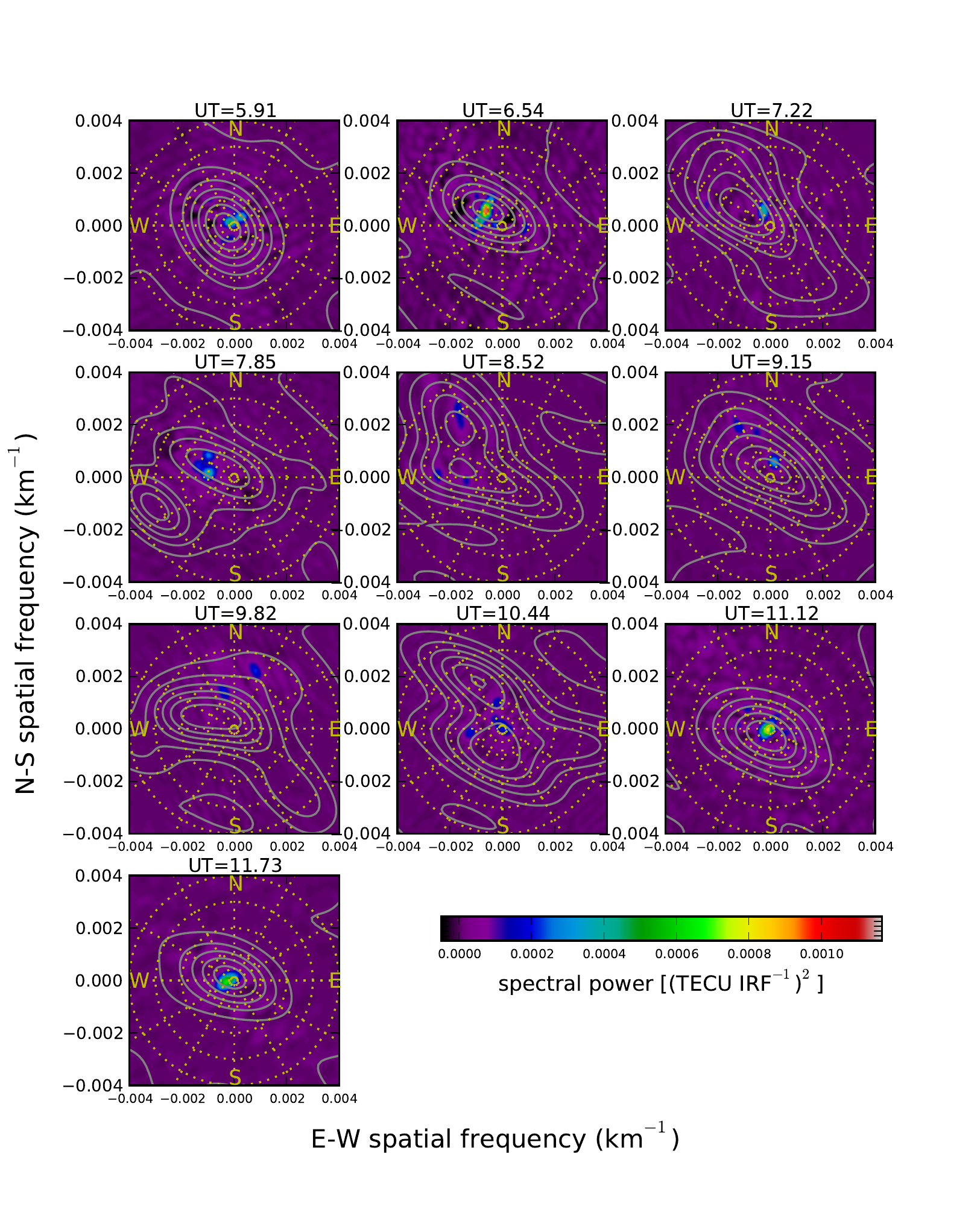}
\caption{Images of the mean spectral power over all temporal frequencies from a combined analysis of all satellite/receiver pairs within each $\sim\!\!35$-minute segment.  The contours depict the mean spectral power from analysis of individual satellites shown as images in Fig.\ \ref{single}.}
\label{all}
\end{figure}

The case study to which we have applied these techniques helps to illustrate the complementary nature of VLA and GPS-based analyses.  VLA observations of cosmic sources in the VHF regime provide a powerful, high-precision probe of the TEC gradient, naturally biassing it toward higher spatial frequency fluctuations.  Conversely, GPS observations, with limited amplitude sensitivity, are chiefly sensitive to medium-to-large-scale disturbances.  In this case, the overlap between the VLA and the GPS array used appears to be in the medium-scale regime at wavelengths of about 100 to a few hundred kilometers.  During these observations, this seems to be especially true for northwestward-directed, medium-scale waves.  Waves with wavelengths $\lesssim\!500$ km traveling toward the northwest were detected within the GPS data during segments centered at 07:13, 08:31, 09:09, and 10:26 UT (00:03, 01:21, 01:59, and 07:16 local time; see Fig.\ \ref{all}).  During those times, the VLA data show evidence for smaller, but still medium-scale, waves (wavelengths $\sim\!\!50$--200 km) traveling toward the northwest as well.\par
\begin{figure}
\noindent\includegraphics[width=6in]{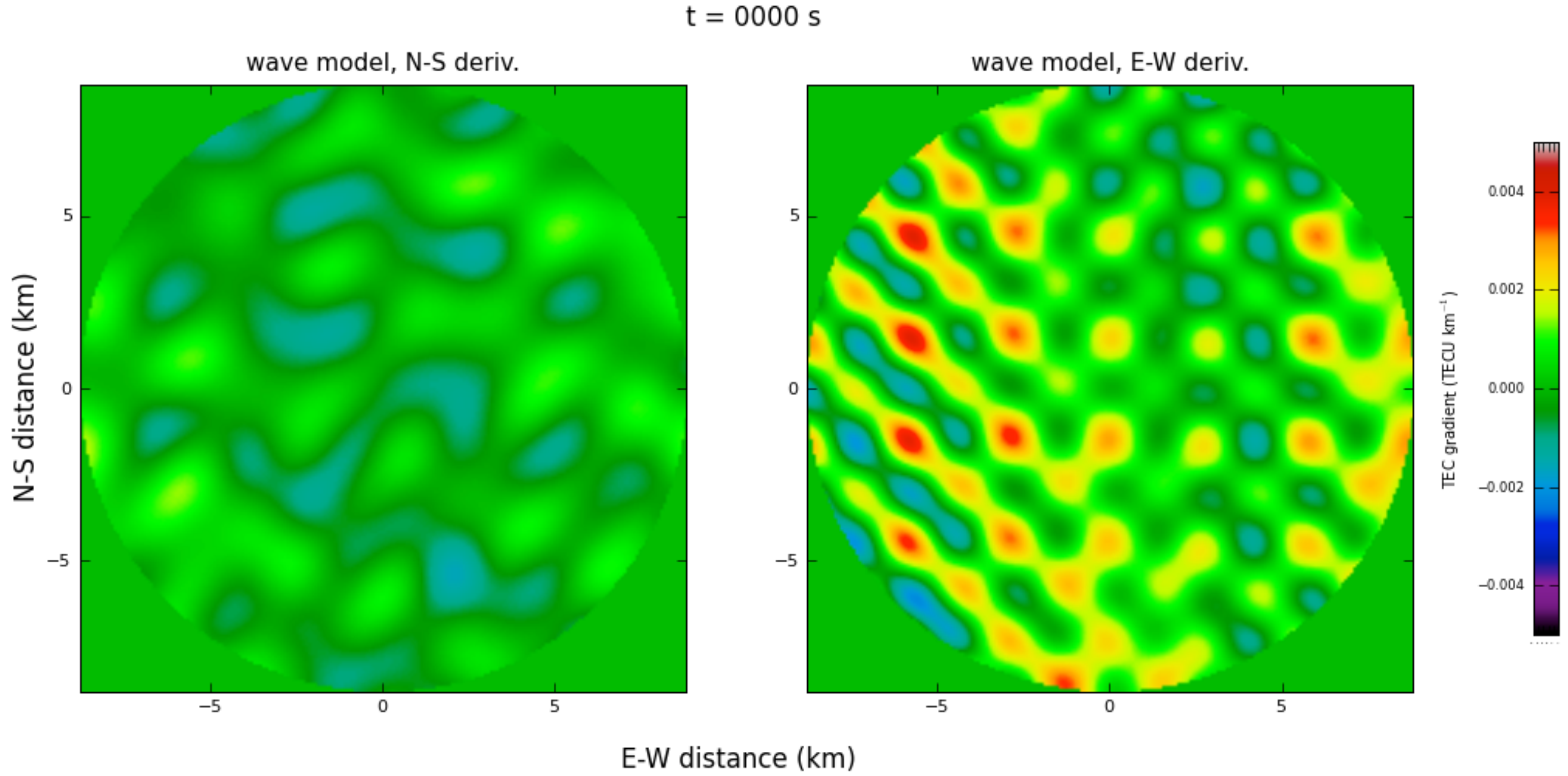}
\caption{The first frame in a movie (available electronically)  depicting the TEC gradient over the VLA P-band circular field of view reconstructed from the spectral analysis for a single $\sim\!\!35$-minute segment (the last one, centered at 11:44 UT; see Sec.\ 4.2 and 4.3).  The north-south component of the gradient is shown in the left panel and the east-west component in the right one.}
\label{vlampg}
\end{figure}
\begin{figure}
\noindent\includegraphics[width=6in]{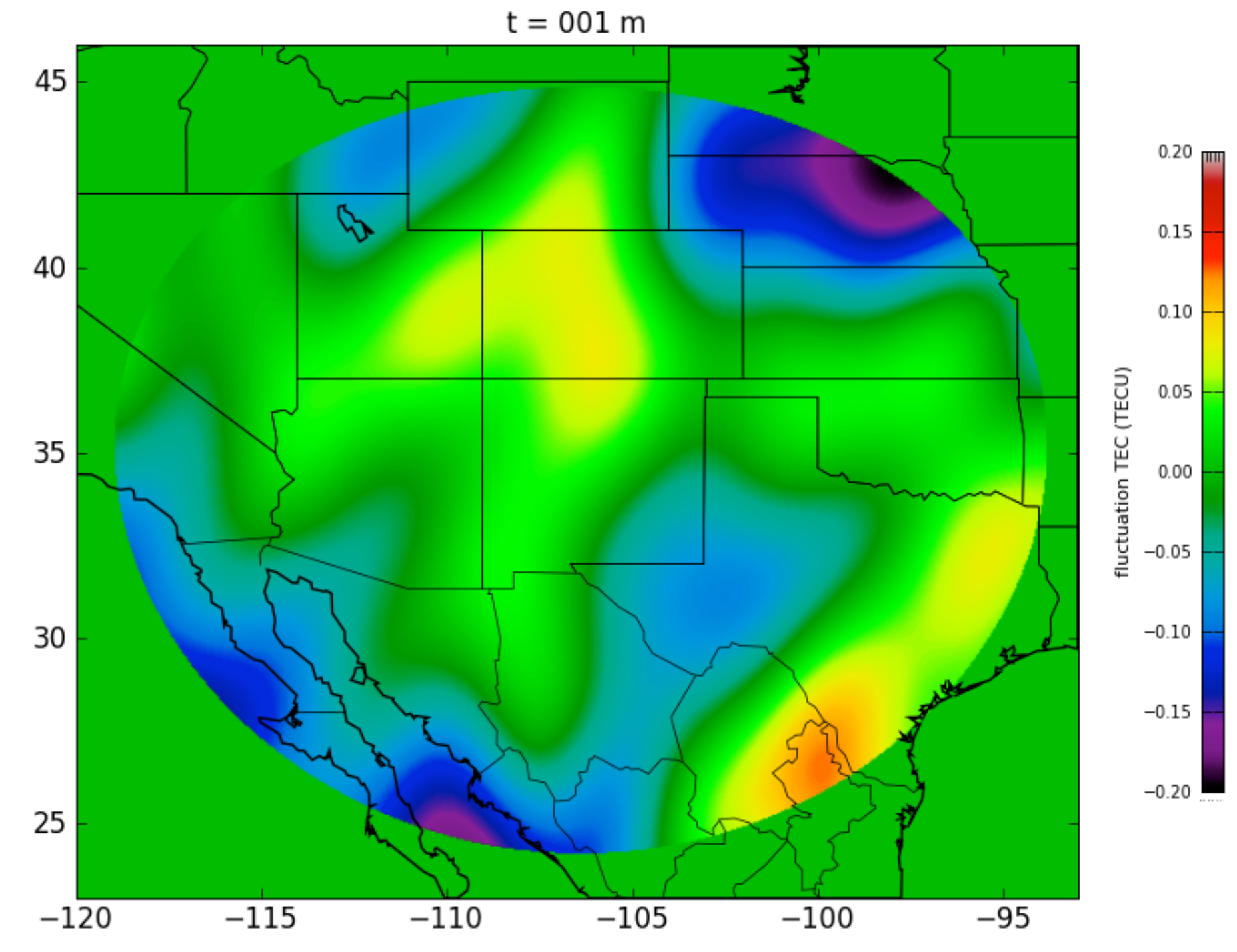}
\caption{The first frame in a movie (available electronically) depicting TEC fluctuations over the approximate field of regard for the New Mexico-based GPS receivers reconstructed from the combined spectral analysis of all satellite/receiver pairs for the entire $\sim\!\!6$-hour observing run (see Sec.\ 4.2).}
\label{gpsmpg}
\end{figure}

For the area near the VLA, such waves are not uncommon during autumn nighttime.  This was demonstrated in a climatological study of TEC gradient fluctuations conducted by \citep{helmboldt12d} using data from a 74 MHz VLA sky survey.  They found that medium-scale waves ($\sim$70--200 km wavelength) traveling in nearly every direction are typically seen during autumn nighttime.  They speculated that these waves are associated with orographic gravity waves generated via wind flow over the relatively mountainous terrain of the southwestern United States and northern Mexico.  The results of this case study are basically consistent with this conclusion.  The relative spatial and temporal complexity of the composite waveforms shown in Fig.\ \ref{vlampg} and \ref{gpsmpg} are consistent with non-linear interactions among such gravity waves that cause them to chaotically break at high altitudes \citep[see, e.g.][]{holton82}.\par
It is interesting to note that while the large-scale waves detected with the GPS data are consistent with the size and speed of typical LSTIDs \citep{borries09}, in this case they seem to be a much more subtle phenomenon, having TEC amplitudes of, at most, $\sim\!\!0.1$ TECU.  In addition, it is extremely unlikely that these LSTIDs, like those described by \citet{borries09}, are associated with either geomagnetic storms or generically increased geomagnetic activity as the $K_p$ and $AE$ indices during the observing run were $\le\!\!3$ and $\le\!\!50$ nT, respectively.  The large-scale waves detected within our GPS data were seen traveling in a variety of directions, but were mostly directed to the west and/or north.  It may be that these LSTIDs represent the low-spatial-frequency portion of a spectrum of waves linked to orographic gravity waves.\par
We also note that often times in the northern hemisphere at night, especially in summer, northwest-to-southeast aligned wavefronts are observed propagating toward either the southwest \citep{hernandez06,tsugawa07,helmboldt12e} or the northeast \citep{shiokawa08,helmboldt12b,helmboldt12e}.  The results presented here show evidence for such waves only in the time interval centered on 07:51 UT.  The VLA data indicate the presence of a relatively prominent northeastward-propagating wave with a $\sim$100-km wavelength and a speed of $\sim$50 m s$^{-1}$.  During this same time interval, the spectrum derived from observations toward a single satellite, G30, visible to the south at at low elevation (see Sec.\ 4.2), has a detection of a southwestward-moving wave with a wavelength of $\sim$300 km.\par
It has been proposed that the origin of these waves is a coupled instability between sporadic-E ($E_s$) layers and the Perkins instability in the $F$-region, which favors disturbances aligned from northwest to southeast \citep{cosgrove04}.  In fact, \citet{helmboldt12e} used a combination of VLA observations and ionosondes data to show that during a particular summer nighttime observing campaign, such wavefronts were only seen when $E_s$ was present.  They and \citet{shiokawa08} also found that the propagation direction for these wavefronts tended to be toward the northeast when the $F$-region was descending, likely due to abrupt shifts in the direction of the $F$-region neutral wind toward the northwest.  During the observations presented here, data from the digisonde station in Boulder, Colorado obtained from the Digital Ionogram Database \citep[DIDBase;][]{reinisch04} showed evidence for $E_s$, but only between 08:00 and 09:00 UT.  The digisonde in Point Arguello, California only indicated the presence of $E_s$ between 11:00 and 11:15 UT.\par
The fact that the only northwest-to-southeast aligned wavefronts detected within this case study were found close to the period of $E_s$ activity found within the Boulder digisonde data is consistent with the hypothesis that $E$-$F$ coupling drives the formation of these nighttime waves.  The fact that the wave detected from the VLA data propagated in the opposite direction from the wave found toward the G30 satellite may suggest a latitudinal dependence in the direction of the $F$-region neutral wind.  In fact, the northeastward-propagating waves detected in Japan have only been detected north of a geomagnetic latitude of about $50^\circ$; the geomagnetic latitude of the VLA is about $42^\circ$.  The results presented here are basically consistent with this as the VLA wave was detected toward A2256 which was in the northern sky at about $30^\circ$ elevation and the GPS wave was detected toward satellite G30, about $13^\circ$ above the southern horizon.  However, in order to explain the lack of detections toward all other GPS satellite visible during the same time interval, the $E$-$F$ coupling-generated waves would have to be relatively weak and/or spotty in nature.  The latter is plausible given the known patchy quality of $E_s$ \citep[e.g.,][]{bernhardt02}.\par
The origin(s) of the phase distortions detected (and corrected for) within our deconvolution method remain(s) unclear.  Instrumental delay errors due to internal clock/hardware limitations \citep[$\sim$1--5 ns; e.g.,][]{coco91} are likely negligible given the comparatively coarse sample intervals of the time series for the TEC gradient (8 s) and TEC (30 s) measured using the VLA and GPS data, respectively.  It is more likely that these phase errors represent a combination of real wavefront distortions and measurement errors, especially within the de-trending process.  Both the $\delta\mbox{TEC}$ (VLA) and TEC (GPS) time series are de-trended to remove slowly varying components resulting from instrumental biases and the effects of the changing receiver/antenna line of sight.  To first order, a simple geometric correction can be applied to remove these line-of-sight effects (i.e., slant-to-vertical TEC corrections).  However, when characterizing wavelike phenomena, higher-order effects, such as the (unknown) three-dimensional orientation(s) of the observed wavefront(s) with respect to the array are much more important.  Therefore, it is expected that any reasonable de-trending scheme will not completely remove these effects, resulting in phase errors in the measured complex quantities, $T(\nu)$, used to generate the power spectrum cubes.\par
It is also likely that there are amplitude errors that were not solved for within the deconvolution process used here and that significantly impacted the results.  In fact, one can see several significant, negative depressions in the spectra shown for the multi-satellite GPS data in Fig.\ \ref{all} that are likely artifacts, which are typically seen within synthesis images when there are uncorrected amplitude errors present.  These errors may arise from errors in receiver tuning and/or bandpass calibration as the TEC and/or TEC gradient amplitude depends on the assumed observing frequency(ies).  Errors in the de-trending process are also likely to contribute.  For the multi-satellite GPS data, the fact that each satellite causes a different Doppler shift for each wave will also introduce non-physical amplitude variations over the array for a particular wave.  Analysis of the waves detected with the method described in Sec.\ 4.4 indicates that this is not generally true, but may be significant in some cases.\par
Therefore, while it may be tempting to use the gain variations solved for within the deconvolution process to explore localized structures that are distorting the passing waves, the data presented here simply do not provide enough information to do so.  To disentangle instrumental and physical effects will require different observations, such as VLA observations of the same cosmic source concurrently at two separate frequencies (e.g., 74 and 330 MHz), each with its own feed system and receiver.  Any gain variations derived from separate spectral analyses of the data from the two bands that are common to both would point to real physical structures as the cause.

\begin{acknowledgments}
Basic research at the Naval Research Laboratory is supported by 6.1 base funding.  The National Radio Astronomy Observatory is a facility of the National Science Foundation operated under cooperative agreement by Associated Universities, Inc.
\end{acknowledgments}

\newpage
%\bibliography{fftspec_draft_v2}{}
%\bibliographystyle{agu08}

\newpage

\end{article}

\clearpage

\clearpage

\clearpage

\clearpage

\clearpage

\clearpage

\clearpage

\clearpage

\clearpage

\clearpage

\clearpage

\clearpage

\clearpage

\end{document}